\documentclass[a4paper,11pt]{article}
\usepackage{jcappub}
\usepackage{lineno}

\title{Bounds on Gravitational Wave Production from Unitarity in an Early NEC-Violating Model}

\author[a]{Pavel Petrov}
\author[b]{Jianing Wang}

\affiliation[a]{Cosmology, Gravity and Astroparticle Physics Group,
Center for Theoretical Physics of the Universe,
Institute for Basic Science (IBS),
Daejeon 34126, Korea}

\affiliation[b]{Kavli Institute for the Physics and Mathematics of the Universe (WPI),
UTIAS, The University of Tokyo,
Kashiwa, Chiba 277-8583, Japan}

\emailAdd{p669590371@ibs.re.kr}
\emailAdd{jianing.wang@ipmu.jp}

\abstract{We study a cosmological scenario featuring an early phase of null energy condition (NEC) violation. Within this framework, we show that perturbative unitarity bounds place strong constraints on both the amplitude and the spectral tilt of primordial gravitational waves. Our analysis is largely insensitive to the detailed realization of the transition between the NEC-violating phase and subsequent cosmological phases, allowing our results to be extended to a broader class of models. Finally, the perturbative unitarity approach employed here is applicable to a wide range of cosmological scenarios.}

\begin{document}
\maketitle
\flushbottom

\section{Introduction}
\label{sec: Introduction}

The recent NANOGrav results~\cite{EPTA:2023fyk, NANOGrav:2023gor, Franciolini:2023pbf, Jiang:2024dxj, Domenech:2024rks, Tan:2024kuk, Kenjale:2024rsc, Kobayashi:2025jkg, donofrio2025validationnanograv15yeardata} have generated considerable interest in the stochastic gravitational-wave (GW) background. A key question is whether such a signal could originate from early-Universe cosmology. Several works have proposed models capable of producing a detectable stochastic GW background~\cite{Tahara:2020fmn,Pi:2022zxs, Borah:2023sbc, Zhu:2023lbf, Vagnozzi:2023lwo, Ye:2023tpz, Chen:2024jca,Pi:2024ert, Akama:2024vgu,sasaki2025unveilingprimordialblackhole}, some of which require a violation of the null energy condition (NEC); for a review, see Ref.~\cite{Rubakov:2014jja}.  

NEC violation can be realized in modified gravity frameworks such as Horndeski gravity~\cite{Horndeski:1974wa}, beyond-Horndeski theories~\cite{Zumalacarregui:2013pma, Gleyzes:2014dya}, or DHOST theories~\cite{Langlois:2015cwa}, all of which admit cosmological solutions with stable NEC violation~\cite{Rubakov:2014jja, Kobayashi:2019hrl}. NEC-violating models have also been employed to construct non-standard early-Universe cosmologies, including Genesis scenarios~\cite{Creminelli:2010ba, Creminelli:2012my, Hinterbichler:2012fr, Elder:2013gya, Pirtskhalava:2014esa, Nishi:2015pta, Kobayashi:2015gga, Ilyas:2020zcb, Zhu:2021ggm} and bouncing models~\cite{Qiu:2011cy, Easson:2011zy, Battarra:2014tga}.  

However, such models may become strongly coupled, rendering classical and perturbative analyses unreliable~\cite{Nicolis:2009qm, Ageeva:2018lko, Ageeva:2020gti, Ageeva:2020buc, Ageeva:2022fyq, Cai:2022ori, Ageeva:2022asq, Ageeva:2023nwf}. In classical analyses, strong coupling arises when cubic terms in the perturbation Lagrangian dominate over quadratic ones, significantly altering the dynamics~\cite{Baumann:2011dt, Bueno:2016xff, Hu:2023juh}. In a quantum field theory context, it is associated with unitarity bounds on scattering amplitudes~\cite{Nicolis:2009qm, Cannone:2014qna, Escriva:2016cwl, deRham:2017avq, Ageeva:2020gti, Kim:2021pbr, Cai:2022ori, Ageeva:2023nwf}. In non-singular Horndeski cosmologies, such bounds can strongly constrain model parameters~\cite{Ageeva:2020gti, Cai:2022ori} or even rule out specific constructions~\cite{Ageeva:2022asq, GilChoi:2025hbs}. Therefore, it is important to examine whether these unitarity bounds would constrain the amplitude and spectral tilt of primordial gravitational waves.  

To illustrate our approach, we analyze a specific model~\cite{Tahara:2020fmn} and explicitly evaluate the corresponding unitarity bounds. We find that perturbative unitarity strongly constrains the production of primordial GWs during the early NEC-violating phase. 
In particular, we derive constraints on the model parameters from perturbative unitarity. We then numerically evaluate the corresponding bounds at the NANOGrav reference frequency and determine the allowed parameter space for $(A_T,\,n_T)$. Within this model, achieving a blue-tilted signal with a sufficiently large amplitude is highly constrained. Importantly, these results do not depend on the specific realization of the transition from the NEC-violating phase to inflation, making them applicable to a broader class of early-Universe scenarios.

This paper is organized as follows. Section~\ref{sec: The model} describes the model. Section~\ref{sec: The estimating of the strong coupling scale} estimates the strong-coupling scale using dimensional analysis and identifies the most relevant cubic terms. Section~\ref{sec: Exact Unitarity Bounds} derives constraints from perturbative unitarity. We conclude in Section~\ref{sec: Conclusion}. Appendix~\ref{app: The cubic action for all sectors} presents the cubic actions for all sectors, Appendix~\ref{app: Dimensional analysis} details the dimensional analysis, and Appendix~\ref{app: Calculating unitary bounds for sub-leading terms} derives the unitarity bounds from scattering amplitudes.

\section{The Model}
\label{sec: The model}

The model considered in Ref.~\cite{Tahara:2020fmn} is constructed within a subclass of Horndeski gravity and is defined by the following covariant Lagrangian:
\begin{equation}
\label{eq: Tahara model covariant Lagrangian}
\mathcal{L} = G_2(\phi, X) - G_3(\phi, X) \square \phi + \frac{M_{\mathrm{Pl}}^2}{2} R\;,
\end{equation}
where $X = -\frac{1}{2} g^{\mu \nu} \partial_\mu \phi \partial_\nu \phi$, $\square \phi = g^{\mu \nu} \nabla_\mu \nabla_\nu \phi$, 
$R$ is the Ricci scalar, $M_{\mathrm{Pl}}$ is the reduced Planck mass. We adopt the metric signature $(-,+,+,+)$. 

At early times (as discussed in Section~\ref{sec: Introduction}, we focus on the NEC-violating stage relevant for the gravitational wave signal), the functions $G_2$ and $G_3$ are chosen as
\begin{equation}
    G_2(\phi, X) = \beta X - V(\phi)\;,\quad 
    G_3(\phi, X) = \frac{\alpha M_{\mathrm{Pl}}}{U(\phi)} X\;,
\end{equation}
where the potential $V(\phi)$ coincides with the function $U(\phi)$, given by
\begin{align}
\label{eq: formOfUandV}
V(\phi) = U(\phi) = U_0 e^{2\phi / M_{\mathrm{Pl}}}, \quad U_0 = \text{const} > 0.
\end{align}
Here, $\alpha$ and $\beta$ are dimensionless constants.
The model~\eqref{eq: Tahara model covariant Lagrangian} admits a power-law expansion of the Universe
\begin{align}
\label{eq: Tahara Model solution}
    a &= a_0 (-t)^{-p}, \quad H = \frac{p}{-t}, \quad t < 0,\; p > 0\;, \\ \nonumber
    \frac{\phi}{M_{\mathrm{Pl}}} &= \ln \left[\frac{M_{\mathrm{Pl}}}{\sqrt{q U_0}(-t)}\right]\;,
\end{align}
where $p$ and $q$ are also dimensionless parameters. Their relation to $\alpha$ and $\beta$ are given by
\begin{equation}
    \alpha = \frac{2p}{q} - \frac{2}{q^2(1 + 3p)}\;, \quad 
    \beta  = -2p(2 + 3p) + \frac{2}{q}\;.
\end{equation}
This model predicts a blue-tilted tensor spectrum with spectral index
\begin{align}
\label{eq:nT value}
    n_T = \frac{2}{p+1}>0\;.
\end{align}

The behavior of the background solution~\eqref{eq: Tahara Model solution} resembles the Modified Genesis scenario introduced in~\cite{Libanov:2016kfc}. Specifically, the scale factor and Hubble parameter asymptotically vanish in the infinite past. Let us now discuss the geodesic (in)completeness of this solution.
The condition for geodesic past completeness in FLRW cosmology is given by~\cite{Libanov:2016kfc, Creminelli:2016zwa}
\begin{align*}
    \int^{T}_{-\infty} \mathrm{d}t\, a(t) = \infty\;,
\end{align*}
where $T$ is some arbitrary finite time.
This condition can also be interpreted as geodesic completeness for gravitons~\cite{Creminelli:2016zwa, Kobayashi:2019hrl}. For solution~\eqref{eq: Tahara Model solution}, this condition implies $p \leq 1$. 
However, to remain within the Horndeski class while ensuring the stability of the background throughout the entire evolution, one must satisfy the conditions indicated in \color{black}{}the No-go theorem~\cite{Kobayashi:2016xpl, Libanov:2016kfc}. This theorem states that, for a fully stable cosmology within Horndeski gravity, the following conditions must hold:
\begin{align*}
    \int_{-\infty}^{T} a(t)(\mathcal{F}_T + \mathcal{F}_S) \, \mathrm{d}t < \infty\;, \qquad 
    \int_{T}^{+\infty} a(t)(\mathcal{F}_T + \mathcal{F}_S) \, \mathrm{d}t < \infty\;,
\end{align*}
where $\mathcal{F}_T$ and $\mathcal{F}_S$ are the gradient coefficients for tensor and scalar perturbations, respectively. In the Einstein frame, these conditions directly contradict geodesic completeness for graviton propagation.
We therefore make two important observations. First, the current model is geodesically incomplete in the past. Second, due to the No-go theorem~\cite{Kobayashi:2016xpl, Libanov:2016kfc}, one must consider only the case $p > 1$ to ensure a stable cosmological solution. 

From here and below we will choose unitary gauge ($\delta\phi = 0,$ $\phi = \phi(t)$) and define the cosmological perturbations as follows, for example see Ref.~\cite{Kobayashi:2011nu}: 
\begin{equation}
\mathrm{d} s^2=-N^2 \mathrm{~d} t^2+\gamma_{i j}\left(\mathrm{d} x^i+N^i \mathrm{~d} t\right)\left(\mathrm{d} x^j+N^j \mathrm{~d} t\right),
\end{equation}
where
\begin{equation}
N= 1+\widetilde{\alpha}, \quad N_i=\partial_i \widetilde{\beta}, \quad \gamma_{i j}=a^2(t) e^{2 \mathcal{R}}\left(\mathrm{e}^h\right)_{i j},
\end{equation}
where $\widetilde{\alpha}$ and $\widetilde{\beta}$ are non-dynamical constraints, while $\mathcal{R}$ is scalar perturbations; $h_{ij}$ is transverse and traceless tensor perturbation. Consequently we have three propagating degrees of freedom.

\section{Estimation of the Strong Coupling Scale}
\label{sec: The estimating of the strong coupling scale}

In Horndeski Genesis models with strong gravity in the past~\cite{Ageeva:2021yik}, the most restrictive condition typically arises from the cubic scalar action~\cite{Ageeva:2020buc}. After performing a conformal transformation to the Jordan frame, the NEC-violating phase in Ref.~\cite{Tahara:2020fmn} closely resembles that of the model in Ref.~\cite{Ageeva:2020buc}, suggesting that the cubic scalar action provides the strongest bounds. Nevertheless, this expectation requires explicit verification, which we leave for future work. Therefore, in this work we focus on the cubic Lagrangian for perturbations.

In this section, we identify the most relevant terms in the cubic actions that contribute to the strong-coupling condition and estimate the corresponding unitarity bounds. All calculations in this section are performed under the following approximations:
\begin{itemize}
    \item We consider each term separately; as a result, our estimates are accurate only up to possible cancellations in the cubic action. 
    \item Tensor structures and all numerical coefficients are omitted.
\end{itemize}
In other words, we retain only the dimensional quantities, and therefore refer to this approach as \textit{dimensional analysis}. This method allows us to identify the cubic terms most relevant for the unitarity bounds. Subsequently, in Subsection~\ref{sec: Most relevant terms for the perturbative unitarity bounds}, we explicitly present these terms and estimate the corresponding bounds.

\subsection{The Method}
\label{subsec: The method}

Since tensor structures and numerical coefficients are neglected, tensor perturbations can be treated as additional scalar fields. Thus, in this subsection, we restrict ourselves to an effective multi-scalar field case. Under these 
assumptions, the quadratic actions can be written schematically as
\begin{align}
\label{eq: structure of quadratic action}
\mathcal{S}_{\mathcal{R}\mathcal{R}}^{(2)} &\propto \int \mathrm{d}t\, \mathrm{d}^{3}x\, a^3
\left[
  M_{\mathrm{Pl}}^2
  \left(\frac{\partial \mathcal{R}}{\partial t}\right)^{2}
  - \frac{M_{\mathrm{Pl}}^2}{a^2} \left(\vec{\nabla} \mathcal{R}\right)^{2}
\right], \\
\mathcal{S}_{hh}^{(2)} &\propto \int \mathrm{d}t\, \mathrm{d}^{3}x\, a^3
\left[
  M_{\mathrm{Pl}}^2
  \left(\frac{\partial h}{\partial t}\right)^2
  - \frac{M_{\mathrm{Pl}}^2}{a^2} h_{,k}h_{,k}
\right], \nonumber
\end{align}
where $\mathcal{R}$ denotes the scalar perturbation in unitary gauge, and $h$ represents the tensor perturbation.  
We introduce new variables $u_{(\mathcal{R})}$, $u_{(h)}$, and the conformal time $\eta$:
\begin{align*}
    u_{(\mathcal{R})} &= M_{\mathrm{Pl}}\, a\, \mathcal{R},\\
    u_{(h)} &= M_{\mathrm{Pl}}\, a\, h,\\
    \eta & \propto \int \frac{\mathrm{d}t}{a} \propto -\frac{1}{a_0}(-t)^{1+p}, \quad -t \propto (-\eta a_0)^{\frac{1}{1+p}},
\end{align*}
where $t$ is the cosmic time  in the Einstein frame. The quadratic actions \eqref{eq: structure of quadratic action} then take the well-known form
\begin{align}
\label{eq: quadratic action-m2eff}
\mathcal{S}_{\mathcal{R}\mathcal{R}}^{(2)} &\propto \int \mathrm{d}\eta\, \mathrm{d}^{3}x
\left[
  \left(\frac{\partial u_{(\mathcal{R})}}{\partial \eta}\right)^{2}
  - \left(\nabla u_{(\mathcal{R})}\right)^{2}
  - m_{\mathrm{eff}}^2 u_{(\mathcal{R})}^2
\right],\\
\mathcal{S}_{hh}^{(2)} &\propto \int \mathrm{d}\eta\, \mathrm{d}^{3}x
\left[
  \left(\frac{\partial u_{(h)}}{\partial \eta}\right)^2
  - u_{(h),k}u_{(h),k}
  - m_{\mathrm{eff}}^2 u_{(h)}^2
\right], \nonumber
\end{align}
where the effective mass is defined as
\begin{align*}
    m_{\mathrm{eff}}^2 \equiv -\frac{a''}{a},
\end{align*}
and primes denote derivatives with respect to conformal time.  

Expression \eqref{eq: quadratic action-m2eff} suggests the existence of an infrared cutoff $E_{\mathrm{IR}}$, estimated as 
\begin{align}
\label{eq: EIR}
    E_{\mathrm{IR}} \propto \max\!\left\{ |m_{\mathrm{eff}}|,\, \left|\frac{ m_{\mathrm{eff}}^{\prime}}{m_{\mathrm{eff}}}\right|\right\}.
\end{align}
We will also refer to \(E_{\mathrm{IR}}\) as the classical energy scale \(E_{\mathrm{class}}\), since this is the energy associated with the dynamics of the classical background.
In our model,
\begin{equation}
a \propto a_0 (-\eta a_0)^{\frac{-p}{1+p}}, \quad m_{\mathrm{eff}}^2 \propto (-\eta)^{-2}, \quad \left|\frac{ m_{\mathrm{eff}}^{\prime}}{m_{\mathrm{eff}}}\right| \propto (-\eta)^{-1}.
\end{equation}
Hence, $|m_{\mathrm{eff}}|$ and $\left| m_{\mathrm{eff}}^{\prime}/ m_{\mathrm{eff}}\right|$ are of the same order, implying $E_{\mathrm{IR}} \propto (-\eta)^{-1}$.

In this article, we focus on the strong-coupling scale, which is a purely UV phenomenon. 
Therefore, we consider only the regime in which the scattering energy $E_{\mathrm{scatter}}$ is much higher than the IR cutoff—i.e., when the scattering time is much shorter than the characteristic timescale of classical evolution.

Now we turn to the interaction terms. The cubic action can be written schematically as
\begin{align}
\label{eq: structure of the cubic Lagrangian}
    \mathcal{S}^{(3)} \propto \int \mathrm{d}\eta\, \mathrm{d}^3x \sum_{\text{all terms}} \widetilde{\Lambda}\, (\partial^n) u_{(\mathcal{R})}^a u_{(h)}^b, \quad a+b=3,
\end{align}
where $\widetilde{\Lambda}$ denotes a slowly-varying coupling constants that generally depends on time. Here, $\partial$ represents a general derivative operator. In Appendix~\ref{app: Dimensional analysis}, we explicitly show that for the model \eqref{eq: Tahara model covariant Lagrangian}, the cubic Lagrangian indeed takes the form \eqref{eq: structure of the cubic Lagrangian}.

In the limit where the scattering energy greatly exceeds the IR cutoff, one can define in/out states and compute the unitarity bounds. For the validity of such computations, we require
\begin{align}
\label{eq: validity of scattering}
    E_{\mathrm{scatter}} &\gg E_{\mathrm{IR}},\\
    E_{\mathrm{scatter}} &\gg \left|\frac{\partial_{\eta}\widetilde{\Lambda}}{\widetilde{\Lambda}}\right|, \quad \text{for all possible }\widetilde{\Lambda}. \nonumber
\end{align}
The second condition ensures that the timescale over which the vertex coefficients vary is much longer than the scattering timescale. Hence, $\widetilde{\Lambda}$ can be treated as slowly-varying constant during the scattering process. 

The matrix element for \(2 \to 2\) scattering is estimated as
\begin{align*}
    \mathcal{M} \propto \frac{1}{E^2} (\widetilde{\Lambda} E^n)^2 = \widetilde{\Lambda}^2 E^{2n-2},
\end{align*}
where $E$ denotes the scattering energy satisfying \eqref{eq: validity of scattering}.  
For the case of non-unity speed of particles, the partial-wave amplitudes (PWA) are defined as~\cite{Ageeva:2022byg}:
\begin{align}\label{def:PWA}
    \widetilde{a}_{\ell}(s)=\frac{1}{64 \pi c_S^3} \int_{-1}^1 \mathrm{d}(\cos \vartheta) P_{\ell}(\cos \vartheta) \mathcal{M}(s, \cos \vartheta) \propto \mathcal{M}, \quad s=\left(p_1+p_2\right)^2.
\end{align}
The unitarity bound, which follows from the optical theorem ($\mathrm{Im}(\widetilde{a}_\ell)=|\widetilde{a}_\ell|^2$), reads
\begin{align*}
    |\mathrm{Re}(\widetilde{a}_\ell)| < \frac{1}{2}.
\end{align*}
The bound is saturated at the \textit{strong-coupling energy} $E_{\mathrm{strong}}$, defined as
\begin{align}\label{def:Estrong}
    |\mathrm{Re}[\widetilde{a}_\ell(E_{\mathrm{strong}}, \eta)]| = \frac{1}{2}.
\end{align}
Hence, the corresponding strong-coupling scale associated with each $\widetilde{\Lambda}$ can be estimated as
\begin{align}
    \label{eq: estimate strong-coupling scale}
    E_{\mathrm{strong}} \propto |\widetilde{\Lambda}|^{\frac{1}{1-n}}.
\end{align}
The validity of the effective field theory then requires
\begin{align*}
    E_{\mathrm{strong}} > E_{\mathrm{scatter}}.
\end{align*}
Combining this with condition \eqref{eq: validity of scattering}, we obtain the criteria for the validity of the classical description:
\begin{align*}
    E_{\mathrm{strong}} \gg E_{\mathrm{IR}}, \quad 
    E_{\mathrm{strong}} \gg \left|\frac{\partial_{\eta}\widetilde{\Lambda}}{\widetilde{\Lambda}}\right|.
\end{align*}
The coefficients $\widetilde{\Lambda}_i$ that yield the lowest strong-coupling energy scale correspond to the \textit{most relevant terms}. These are presented in Subsection~\ref{sec: Most relevant terms for the perturbative unitarity bounds}, while the supplementary calculations can be found in Appendix~\ref{app: Dimensional analysis}. In Section~\ref{sec: Exact Unitarity Bounds}, we consider only these most relevant terms and compute the exact unitarity bounds, including all numerical coefficients and tensor structures.

\subsection{Most Relevant Terms}
\label{sec: Most relevant terms for the perturbative unitarity bounds}

The most stringent unitarity bounds originate from the following interaction terms (see Appendix~\ref{app: Dimensional analysis} for detailed calculations):
\begin{align}
\label{eq: L}
    S^{L} = \int \mathrm{~d} \eta \mathrm{~d}^3 x \frac{1}{2}\left\{ 2~\widetilde{ \Lambda}_{7,(1)} \left[\mathcal{R}_c^{\prime}\left(\partial^2 \mathcal{R}_c\right)^2 - \mathcal{R}_c^{\prime}\left(\partial_i \partial_j \mathcal{R}_c\right)^2 \right]\right\}\;,\;\;E^{L}_{\mathrm{str}} \sim a_0 M_{\mathrm{Pl}}^{1/4} H^{3/4 + p}\;.
\end{align}
The next-to-leading order contribution arises from: 
\begin{equation}
\begin{aligned}
\label{eq: SL}
S^{SL}=\int \mathrm{~d} \eta \mathrm{~d}^3 x~\frac{1}{2}&\left\{ \left(2\widetilde{\Lambda}_{3,(1)}\right) \left[\left(\mathcal{R}_c^{\prime}\right)^2 \partial^2 \mathcal{R}_c\right] \right.\\
+&\left(2\widetilde{\Lambda}_{7,(2)}-2\widetilde{\Lambda}_{8}\right) \left[-\mathcal{R}_c\left(\partial^2 \mathcal{R}_c\right)^2 + \mathcal{R}_c\left(\partial_i \partial_j \mathcal{R}_c\right)^2\right]\\
+& \left(2\widetilde{\Lambda}_{9}\right) \left[ \partial^2 \mathcal{R}_c\left(\partial_i \mathcal{R}_c\right)^2\right]\\
+&\left. \left(2\widetilde{\Lambda}_{16,(1)}\right) \left[ \mathcal{R}_c^{\prime}\left(\partial_i \partial_j \mathcal{R}_c\right) \partial_i \partial_j \psi_a\right]\right\}\;,\;\;E^{SL}_{\mathrm{str}} \sim a_0 M_{\mathrm{Pl}}^{1/3} H^{2/3 + p}\;.
\end{aligned}
\end{equation}
Here, the subscript $L$ denotes the leading-order contribution, while $SL$ denotes the next-to-leading-order contribution.

Now let us concentrate on \( S^L \) and estimate the constraints that arise from the unitarity bounds. From  ~\eqref{eq: EIR}, the classical energy scale is estimated as
\[
    E_{\text{class}} \sim a_0\, H^{p+1}\;.
\]
The condition for avoiding unitarity violation is therefore
\begin{equation}\label{eq:estimate}
    \frac{E_{\text{str}}^{L}}{E_{\text{class}}} \gg 1 
    \quad \Rightarrow \quad  
    \left(\frac{H}{M_{\text{Pl}}}\right)^{-1/4} \gg 1\;.
\end{equation}
If one assumes a rapid transition from the NEC-violating phase to the subsequent inflationary phase, the maximum value of the Hubble parameter during the early NEC-violating stage is related to the number of inflationary e-folds \(N\) as (see Ref.~\cite{Tahara:2020fmn})
\begin{equation}\label{eq:Hmax}
    H_{\max} = p \sqrt{\frac{q}{2}}\, m\, (4N)^{3/4}.
\end{equation}
In this work, we adopt the value of the Hubble parameter corresponding to the end of the NEC-violating phase. The Hubble parameter may, however, take larger values during the transition itself and, in some models, can even reach \(\mathcal{O}(0.1)\,M_{\mathrm{Pl}}\)~\cite{Tahara:2020fmn}. The details of this transition are highly model-dependent. Therefore, in order to obtain an optimistic yet transition-independent estimation, we consider the Hubble parameter at the end of the NEC-violating phase, which is typically smaller than its value during the transition.
Here, the parameter \(m\) denotes the effective mass of the inflaton field. Larger values of \(m\) correspond to higher values of the Hubble parameter at the onset of inflation. For further details, including the full Lagrangian describing the NEC-violating phase, the inflationary phase, and the reheating phase, we refer the reader to Ref.~\cite{Tahara:2020fmn}.

Taking representative parameter values  
\[
   p = \frac{3}{2}, \qquad q = 10,\qquad m = 10^{-3} M_{\text{Pl}}, \qquad N= 39.7
\]
as in Ref.~\cite{Tahara:2020fmn}, one finds
\begin{align}  
\label{eq: naive ratio}
\left(\frac{H_{\max}}{M_{\text{Pl}}}\right)^{-1/4} \sim 1.6\;.
\end{align}
Here, $N$ denotes the number of e-folds during the inflationary stage, which may be significantly smaller than the canonical $50$--$60$ e-folds. Nevertheless, the total number of e-folds, including both the pre-inflationary and inflationary stages, should be approximately $50$--$60$ in order to provide a consistent replacement for standard inflation \cite{Powell:2006yg,Bahrami:2015bva,Zhu:2017jew}.

The ratio \eqref{eq: naive ratio} is of order unity, which implies two things. First, this particular choice of the parameter \( m \) is naively disfavored by the unitarity bound. Second, the fact that the ratio is not parametrically large indicates that unitarity bounds can indeed place meaningful constraints on the model parameters. Therefore, it is worthwhile to proceed with an exact calculation of the unitarity bounds for this model. We perform such a calculation in the next section.

Now let us comment on the choice of the cubic action for perturbations. We adopt the action from Ref.~\cite{Ageeva:2020gti}, which is based on Refs.~\cite{Gao:2011qe,Gao:2011vs,DeFelice:2010nf}.  The scalar-sector cubic action in Ref.~\cite{Ageeva:2020gti} explicitly contains terms with up to five derivatives. However, as shown in Ref.~\cite{DeFelice:2011uc}, all five-derivative terms in the cubic action can be eliminated by a sequence of integrations by parts. For this reason, we expect that, after calculating the matrix element corresponding to~\eqref{eq: L}, it should vanish identically. This provides an additional consistency check of our formalism. Indeed, in Section~\ref{subsec: Matrix Element for the Leading-Order Action}, we find precisely this expected behavior.

Moreover, Ref.~\cite{Renaux-Petel:2011zgy} demonstrates that, by using the linear equations of motion, one can apparently eliminate all cubic terms with four derivatives, leaving only operators with two derivatives. It is well known that applying lower-order equations of motion to remove higher-dimensional operators in the Lagrangian is equivalent to performing a field redefinition. A local and invertible field redefinition does not change on-shell $S$-matrix elements, although it can modify off-shell Green's functions and the apparent form of the Lagrangian. 
While such field redefinitions do not affect on-shell observables, they can obscure manifest power counting at the level of the Lagrangian unless all induced higher-order operators are consistently included. We therefore do not expect the power counting introduced in Ref.~\cite{Ageeva:2020buc} to remain valid after such a field redefinition. Indeed, performing a field redefinition (equivalently, using the linear equations of motion) generally generates new higher-order interaction terms. In principle, these induced terms must be taken into account when estimating the strong-coupling scale. For this reason, we refrain from using the simplified cubic action of Ref.~\cite{Renaux-Petel:2011zgy} in our analysis. However, if one is interested solely in computing on-shell three-point correlation functions, the cubic actions of Refs.~\cite{Renaux-Petel:2011zgy,DeFelice:2011uc} do indeed yield the same result.

\section{Unitarity Bounds from the Optical Theorem}
\label{sec: Exact Unitarity Bounds}

To compute the exact perturbative unitarity bounds, we evaluate the matrix elements, extract the partial-wave amplitudes, and finally apply the optical theorem.

We work in the center-of-mass (CoM) frame. The canonical quadratic action for  scalar perturbation is
\begin{equation}
\begin{aligned}\label{eq:canoscalar}
\mathcal{S}_{\mathcal{R}\mathcal{R}}^{(2)} 
&= \frac{1}{2} \int \mathrm{d}\eta \, \mathrm{d}^3x 
\left[ \mathcal{R}_c^{\prime\, 2} - c_S^2 \mathcal{R}_{c,i}^2 - m_{\text{eff}}^2 \mathcal{R}_c^2 \right], \quad c_S^2 \equiv \mathcal{F}_S / \mathcal{G}_S \;.
\end{aligned}
\end{equation}
In the high-energy limit \( k \gg |m_{\text{eff}}| \), the CoM kinematics and dispersion relations are
\begin{equation}
\begin{aligned}
& \vec{p}_1 + \vec{p}_2 = \vec{p}_3 + \vec{p}_4 = \vec{0}, 
\qquad E_1 + E_2 = E_3 + E_4 = E , \\
& |\vec{p}_1| = |\vec{p}_2|, 
\qquad |\vec{p}_3| = |\vec{p}_4|, \qquad  \langle \vec{p}_1, \vec{p}_3\rangle =\cos\vartheta \equiv x,\\
& E_{1,2,3,4} = E/2,
\qquad E_{1,2,3,4} = c_S |\vec{p}_{1,2,3,4}| .
\end{aligned}
\end{equation}
For convenience we define the momentum transfers as
\[
\vec{p}_t \equiv \vec{p}_1 - \vec{p}_3,
\qquad
\vec{p}_u \equiv \vec{p}_1 + \vec{p}_3 .
\]

\subsection{Matrix Element for the Leading-Order Action}
\label{subsec: Matrix Element for the Leading-Order Action}

We now compute the vertices for the three channels of leading interaction \eqref{eq: L}.

\paragraph{(1) \(s\)-channel (two incoming legs and outgoing propagator).}With 
\(
p_1 + p_2 = (E_1, \vec{p}_1) + (E_2, \vec{p}_2) = (E,\vec{0})
\),
the vertex evaluates to
\begin{equation}
\begin{aligned}
& (i \vec{p}_1)^2 (i \vec{p}_2)^2 (iE)
-
(i p_{1,i})(i p_{1,j})
(i p_{2,i})(i p_{2,j})
(iE)
= 0 .
\end{aligned}
\end{equation}

\paragraph{(2) \(t\)-channel (one incoming leg, one outgoing leg and outgoing propagator).}
With
\(
p_1 - p_3 = (0,\vec{p}_t)
\),
\begin{equation}
\begin{aligned}
& (i \vec{p}_1)^2 (-i \vec{p}_t)^2 (iE_3)
+
(-i \vec{p}_3)^2 (-i \vec{p}_t)^2 (-iE_1)
\\
&\quad
-
(i p_{1,i})(i p_{1,j})
(-i p_{t,i})(-i p_{t,j})
(iE_3)
\\
&\quad
-
(-i p_{3,i})(-i p_{3,j})
(-i p_{t,i})(-i p_{t,j})
(-iE_1)
= 0 .
\end{aligned}
\end{equation}

\paragraph{(3) \(u\)-channel (one incoming leg, one outgoing leg and outgoing propagator).}
With
\(
p_1 - p_4 = (0,\vec{p}_u)
\),
\begin{equation}
\begin{aligned}
& (i \vec{p}_1)^2 (-i \vec{p}_u)^2 (iE_4)
+
(-i \vec{p}_4)^2 (-i \vec{p}_u)^2 (-iE_1)
\\
&\quad
-
(i p_{1,i})(i p_{1,j})
(-i p_{u,i})(-i p_{u,j})
(iE_4)
\\
&\quad
-
(-i p_{4,i})(-i p_{4,j})
(-i p_{u,i})(-i p_{u,j})
(-iE_1)
= 0 .
\end{aligned}
\end{equation}
Thus, the full tree-level matrix element arising from the leading interactions cancels identically. This verifies the statement made in Section~\ref{sec: Most relevant terms for the perturbative unitarity bounds}. Consequently, these operators do not generate a strong-coupling scale. Therefore, to estimate the unitarity bounds, we analyze the sub-leading interactions~\eqref{eq: SL} in the next subsection.

One of the main issues that may arise when analyzing unitarity in the unitary gauge is that the apparent cutoff could be a gauge artifact rather than the true physical cutoff of the theory; see, for example, Refs.~\cite{Grosse-Knetter:1992tbp,Burgess:2010zq,Karananas:2022byw,Giudice:2024tcp}. However, we do not expect the cutoff found in this paper to be a gauge artifact, for two reasons.

First, the decoupling limit ($H \to 0$, $\dot H/H \to 0$) is well defined, and no interaction terms in the cubic Lagrangian diverge in this limit.  

Second, in the language of the EFT of dark energy/inflation~\cite{Cheung:2007st,  Gubitosi:2012hu, Tsujikawa:2014mba}, the operator corresponding to the Kinetic Gravity Braiding (KGB) sector takes the form
\begin{align}
\label{eq:kgb_eft_term}
-\frac{\overline{m}^3(t)}{2}\,\delta g^{00}\,\delta K \, .
\end{align}
$\overline{m}^3(t)$ is braiding coefficient, $\delta g^{00}$ is lapse perturbation, $\delta K$ is extrinsic curvature perturbation.
We expect the most relevant unitarity bounds to arise from the higher-derivative sector, namely from the $G_3\,\Box\phi$ operator. For this reason, in the discussion below we focus on the term~\eqref{eq:kgb_eft_term}.
We then reintroduce the Goldstone boson, which restores diffeomorphism invariance via the St\"uckelberg trick, by performing an infinitesimal time diffeomorphism:
\[
t \to t + \pi(x), \qquad x^i \to x^i \, .
\]
After reintroducing the Goldstone boson, all gauge artifacts should disappear.
The operators $\delta g^{00}$ and $\delta K$ are not invariant under four-dimensional diffeomorphisms and therefore transform as
\begin{align*}
g^{00} &\;\to\; g^{00} + 2 g^{0\mu}\partial_\mu \pi + g^{\mu\nu}\partial_\mu \pi\,\partial_\nu \pi \, , \\
\delta K &\;\to\; \delta K - 3 \dot H\, \pi - a^{-2}\nabla^2 \pi \, .
\end{align*}
It then follows that the operator~\eqref{eq:kgb_eft_term} contributes the following term to the cubic action for the Goldstone field $\pi$ in the decoupling limit:
\begin{align*}
(\nabla^2 \pi)\,(\partial_\mu \pi)^2 \, .
\end{align*}
The term above is a dimension-seven operator containing four derivatives, precisely matching both the dimensionality and the derivative count of the terms in the action~\eqref{eq: SL} used to compute perturbative unitarity bounds. Consequently, our method reproduces the same energy scaling of the cutoff as the analysis based on the decoupling limit of the Goldstone boson $\pi$ action.

As expected, in the decoupling limit of the action for the field $\pi$, no cubic terms of dimension eight appear. This fully coincides with the observation made in Section~\ref{subsec: Matrix Element for the Leading-Order Action}.

\subsection{Constraints on the Model Parameters}
\label{subsec: Unitarity Bounds from Sub-leading Terms and Constraints on the Model Parameters}

\begin{figure}[h!]
    \centering
    \includegraphics[width=0.73\textwidth]{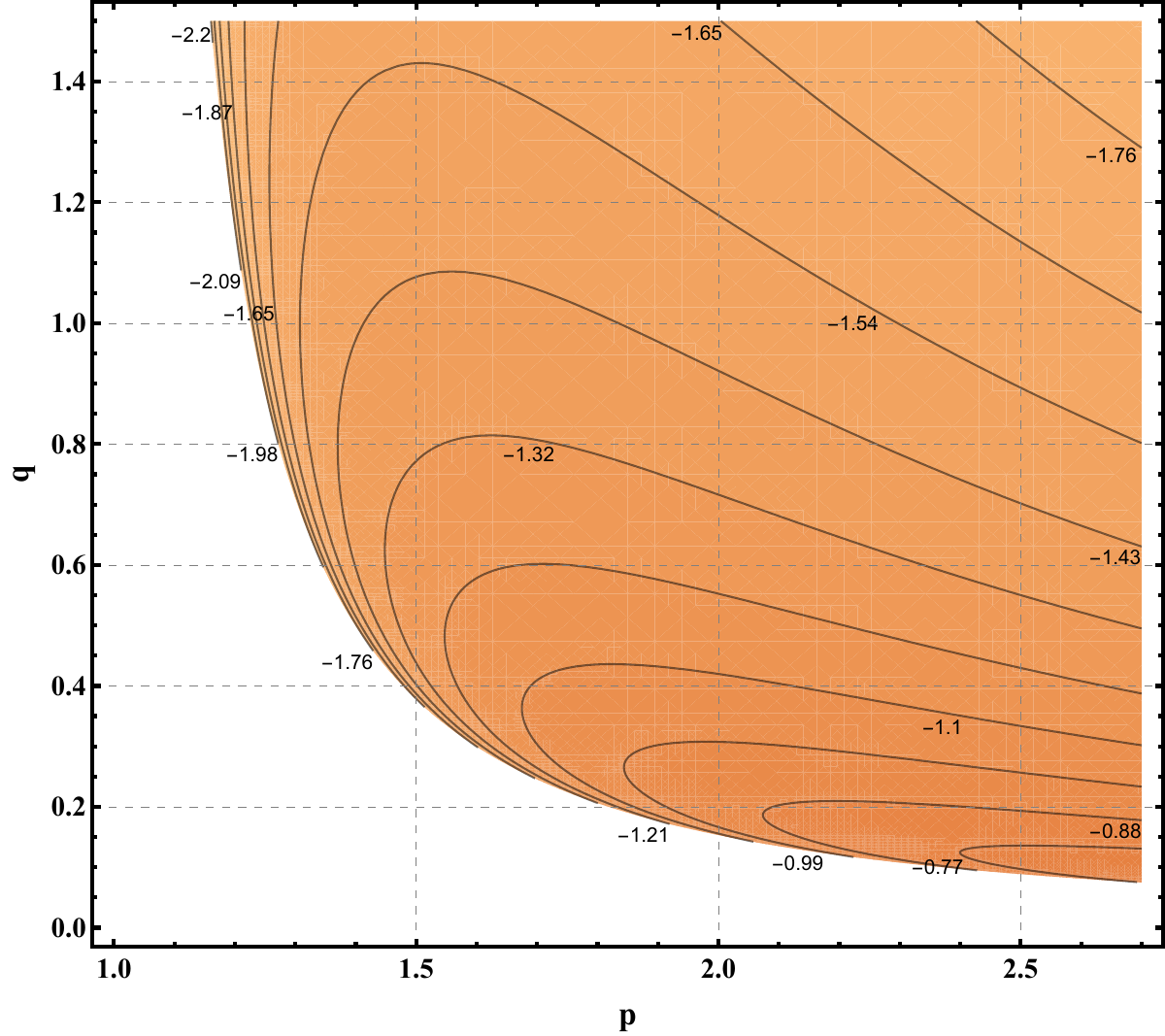}
    \caption{Values of $\log_{10}\mathcal{F}$ over the region of the $(p,q)$ plane that satisfies the conditions of stability and sub-luminal propagation.}
    \label{fig:values_of_F}
\end{figure}

Now we consider the sub-leading operators in~\eqref{eq: SL} and compute the corresponding strong-coupling scale. The detailed derivation is presented in Appendix~\ref{app: Calculating unitary bounds for sub-leading terms}. The resulting strong-coupling energy scale is
\begin{align}
 E_{\mathrm{strong}}
=
M_{\mathrm{Pl}}^{1/3} H^{\,p+2/3}\, 2^{7/3} \pi^{1/6} ~a_0 \left|
\, 
c_S^{-9}\,
g(p,q)\,
\left(2 f_0(c_S,p,q)+\frac{2}{3} f_2(c_S,p,q)\right)
\right|^{-1/6}. 
\end{align}
Using \eqref{eq: EIR}, we actually have
\begin{align*}
    E_{\mathrm{class}} = a_0 H^{p+1} h(p), \quad h(p)\equiv \max\Big[(1+2p)^{1/2}p^{-(p+1/2)},\;\;\left|(1+p) p^{-(p+1)}\right|\Big]\;,
\end{align*} 
hence
\begin{align}
\frac{E_{\mathrm{strong}}}{E_{\mathrm{class}}}
=
\left(\frac{H}{M_{\mathrm{Pl}}}\right)^{-1/3} \, 2^{7/3} \pi^{1/6} ~\left|
\, 
c_S^{-9}\,
h^6(p) g(p,q)\,
\left(2 f_0(c_S,p,q)+\frac{2}{3} f_2(c_S,p,q)\right)
\right|^{-1/6}. 
\end{align}
Parametrically, the strongest bounds are expected to arise at the point where the Hubble parameter reaches its maximum value. This occurs approximately at the end of the NEC - violation phase. Neglecting the short transition stage, this maximum Hubble parameter approximately coincides with its value at the beginning of the inflationary stage. Therefore, using the expression \eqref{eq:Hmax} for the maximum possible value of the Hubble parameter, we find that the strongest perturbative unitarity bounds take the following form:

\begin{align}\label{eq:EstrEclass}
\frac{E_{\mathrm{strong}}}{E_{\mathrm{class}}}
=
N^{-1/4}\left(\frac{m}{M_{\mathrm{Pl}}}\right)^{-1/3} \, \mathcal{F}(p,q),
\end{align}
where
\begin{align}\label{eq:Fpq}
\mathcal{F}(p,q)\equiv 4~ \pi^{1/6} ~\left|
\, 
c_S^{-9}~p^2 q\,
h^6(p) g(p,q)\,
\left(2 f_0(c_S,p,q)+\frac{2}{3} f_2(c_S,p,q)\right)
\right|^{-1/6}. 
\end{align}
We evaluate the function $\log_{10}\mathcal{F}$ over the region in the $(p,q)$ plane that satisfies the conditions of stability (i.e., absence of ghost and gradient instabilities) as well as sub-luminal propagation, and show the results in Figure~\ref{fig:values_of_F}. One can observe that the values of $\mathcal{F}$ in this region can be quite small, even reaching $\mathcal{O}(10^{-2})$.

To safely satisfy the perturbative unitarity bounds, the cutoff of the theory must lie sufficiently above the classical energy scale defined in Eq.~\eqref{eq: EIR}. Specifically, we require the ratio appearing in Eq.~\eqref{eq:EstrEclass} to be sufficiently large. In practice, we impose that the ratio of the UV cutoff $E_{\rm strong}$ to the classical energy scale satisfies
\begin{align}
\label{eq:ratio for cutoff}
\frac{E_{\rm strong}}{E_{\rm class}} \gtrsim \mathcal{O}(10),
\end{align}
i.e., at least one order of magnitude. This choice is conservative and illustrative, and is intended to provide the most optimistic estimate of the allowed parameter space. We note that requiring a larger ratio $E_{\rm strong}/E_{\rm class}$ would lead to significantly stronger constraints on the model parameters and to much lower allowed maximum values of $n_T$.

In particular, perturbative unitarity bounds establish a \emph{lower bound} on the parameter $p$, which in turn implies an \emph{upper bound} on the tensor spectral tilt $n_T$, as given in Eq.~\eqref{eq:nT value}. This statement holds because in our model the relation between $p$ and $n_T$ is monotonic. Moreover, the gradient coefficient $\mathcal{F}_S$, the kinetic term $\mathcal{G}_S$, and the sound speed of scalar perturbations $c_S$ depend only on the parameters $p$ and $q$, which considerably simplifies the analysis.

\begin{figure}[h!]
    \centering
    \includegraphics[width=0.73\textwidth]{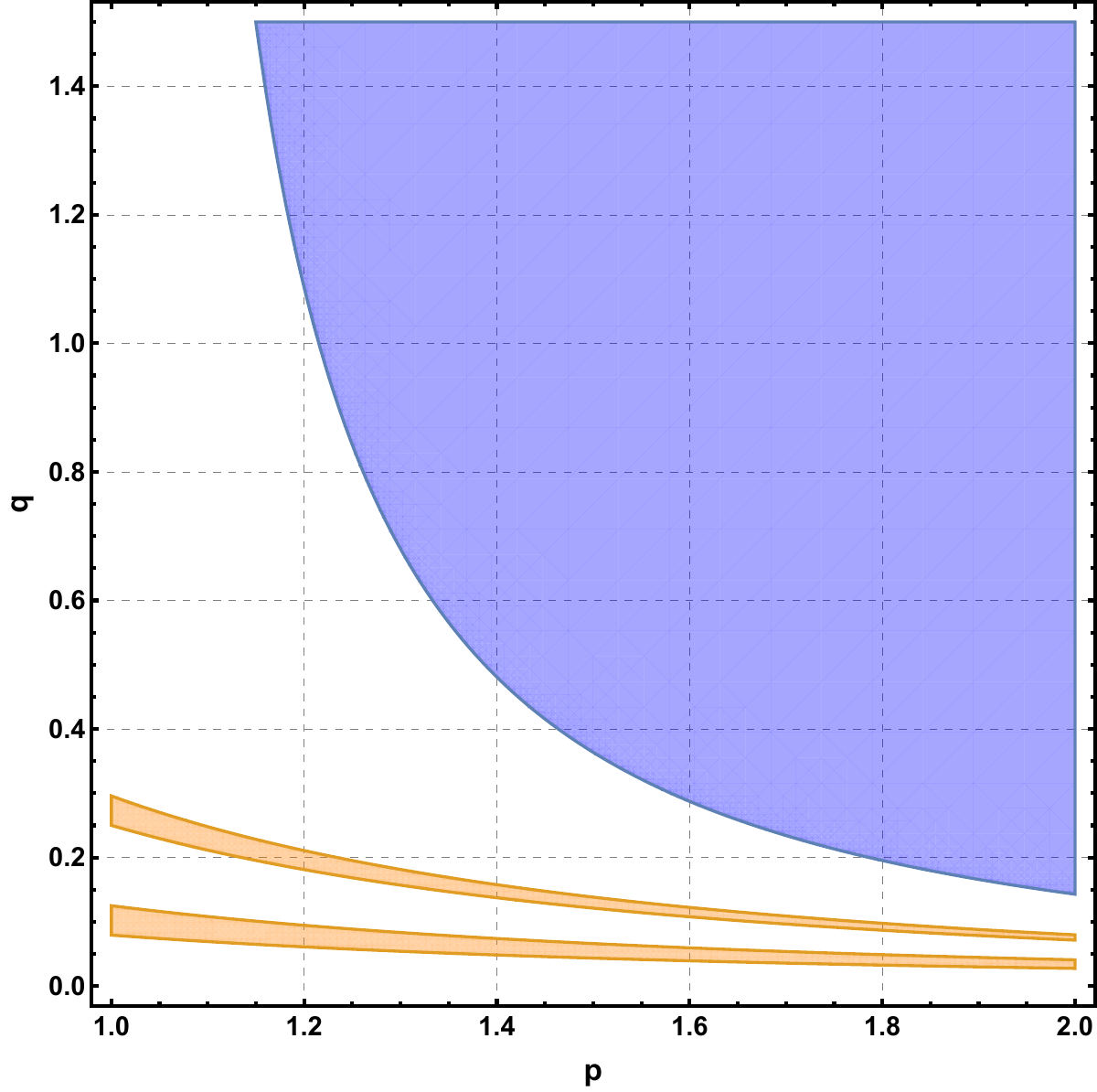}
    \caption{The blue area corresponds to the stability requirements, $\mathcal{F}_S>0$ and $\mathcal{G}_S>0$, while the orange area corresponds to $c_S>1$. }
    \label{fig:stability}
\end{figure}

We can also see from Eq.~\eqref{eq:EstrEclass} that the allowed values of the parameter $m$ are extremely sensitive to the numerical prefactor encoded in the function $\mathcal{F}(p,q)$. In particular, a change of one order of magnitude in $\mathcal{F}$ leads to variations of three orders of magnitude in the bounds on $m$. For this reason, in order to obtain consistent estimates, it is crucial to evaluate this numerical prefactor accurately. This provides an additional motivation for our work, since estimations based solely on power counting or dimensional analysis may lead to misleading conclusions.
\begin{figure}[h!]
    \centering
    \includegraphics[width=0.8\textwidth]{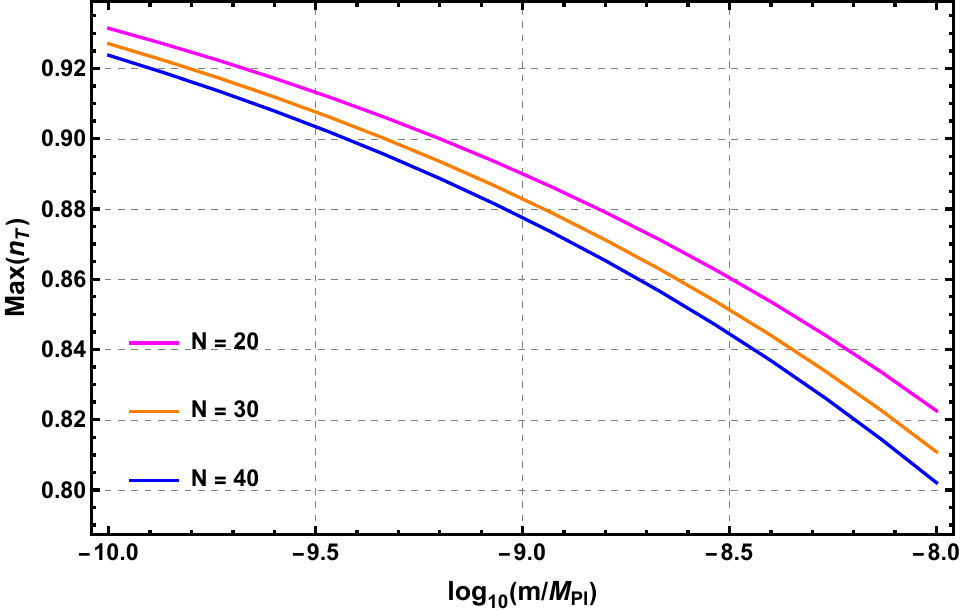}
    \caption{Maximum allowed value of $n_T$ as a function of $\log_{10}(m/M_{\mathrm{Pl}})$ for different $e$-folding numbers $N$ during the inflationary stage. }
    \label{fig:maxNt}
\end{figure}
\begin{figure}[h!]
    \centering
    \includegraphics[width=0.73\textwidth]{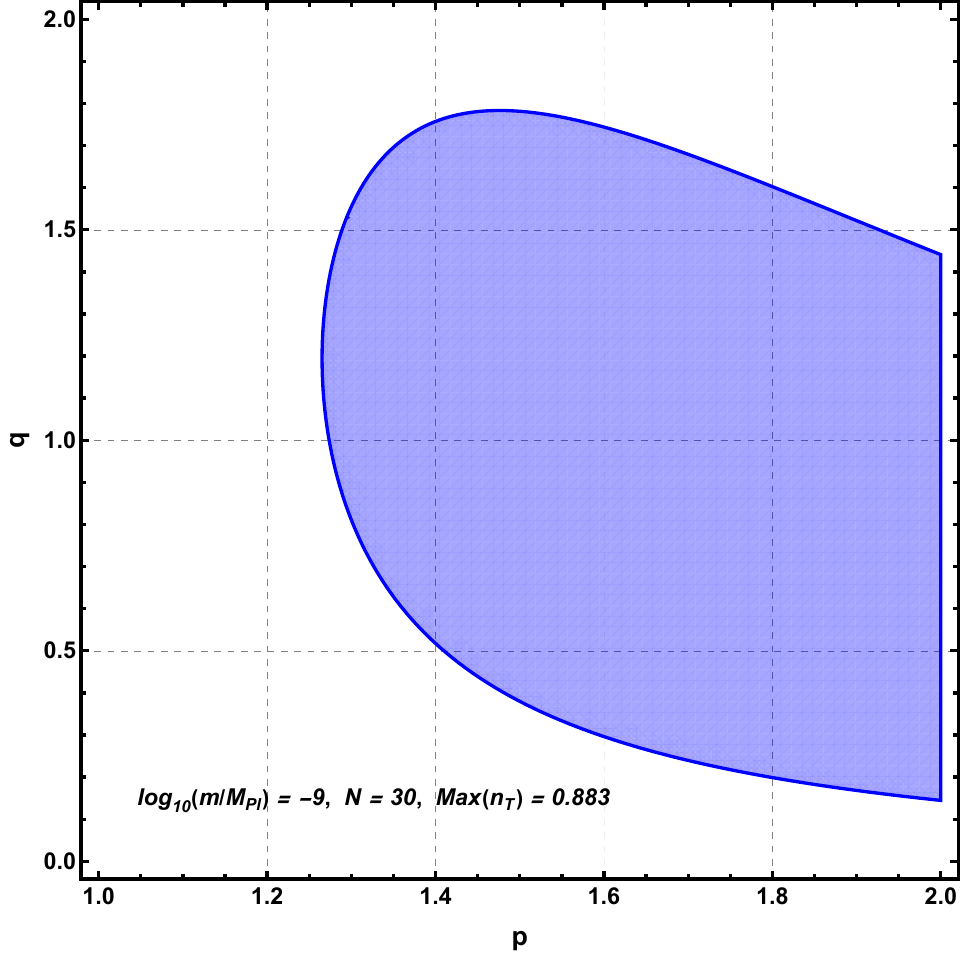}
    \caption{Allowed parameter range in the $(p,q)$ plane for $N = 30$ and $\log_{10}(m/M_{\mathrm{Pl}}) = -9$. }
    \label{fig:pq}
\end{figure}
To illustrate this, we consider different $e$-folding numbers during the inflationary stage and plot the \textit{maximum value of $n_T$ as a function of the model parameter $m$}. For each value of $m$, we vary the parameter $q$ over its allowed range, ensuring that the stability conditions—absence of ghosts and gradient instabilities—are satisfied, and that scalar perturbations propagate sub-luminally ($c_S \leq 1$). We then numerically determine the minimal value of $p$ consistent with these requirements. This procedure ensures that the computed upper bounds on $n_T$ are physically reliable. The results are shown in Figure~\ref{fig:maxNt}. In addition, Figure~\ref{fig:pq} displays the allowed parameter range in the $(p,q)$ plane for $N = 30$ and $\log_{10}(m/M_{\mathrm{Pl}}) = -9$.
The condition forbidding super-luminal propagation is relatively weak and does not affect our conclusions, as shown in Figure~\ref{fig:stability}. Moreover, there is still no clear consensus in the literatures on whether sub-luminal propagation of perturbations in cosmological backgrounds must be strictly imposed, or on how to properly implement causality and positivity bounds in an expanding Universe. For a discussion, see Refs.~\cite{Adams:2006sv, Babichev:2007dw, Hollowood:2015elj, deRham:2020zyh, Creminelli:2022onn, CarrilloGonzalez:2023cbf, Kaplan:2024qtf, CarrilloGonzalez:2023emp, CarrilloGonzalez:2025fqq, Hui:2025aja}.

Now let us comment on the amplitude of primordial gravitational waves. 
Up to numerical factors, the tensor amplitude is proportional to
\begin{align}
    A_T \sim \frac{H_f^2}{M_{\mathrm{Pl}}^2}\;,
\end{align}
where $H_f$ is the value of the Hubble parameter at the time $t_f$ when a given tensor mode freezes out. 
Roughly speaking, perturbative unitarity bounds impose an upper limit on the maximal allowed value of the Hubble parameter. 
Therefore, in order to estimate the \textit{maximal} possible value of $A_T$, it is instructive to consider the limiting case in which the freeze-out time $t_f$ is marginally equal to the time $t_{*}$ at which the Hubble parameter reaches its maximal value~\eqref{eq:Hmax}. 
This occurs at the end of the NEC-violating phase~\eqref{eq:Hmax}.

We then assume instantaneous reheating and take the following values for the present temperature $T_0$ and the reheating temperature $T_{\rm reh}$:
\begin{align}
    T_0 \simeq 10^{-32} M_{\mathrm{Pl}}, 
    \qquad 
    T_{\rm reh} \simeq 10^{-12} M_{\mathrm{Pl}} \;.
\end{align}
The freeze-out time for a given comoving momentum $k_{*}$ is estimated as
\begin{align}
    k_{*} \sim H a \Big|_{t = t_f}\;,
\end{align}
while the scale factor satisfies
\begin{align}
 a(t_f) \,    \frac{a(t_{*})}{a(t_f)} \, e^{N} \, \frac{T_{\rm reh}}{T_0} \simeq 1 \;,
\end{align}
where $t_{*}$ corresponds to the end of the NEC-violating phase, and by convention we set the present-day value of the scale factor to unity. 
Using the assumption $t_f \simeq t_{*}$, one can determine the number of $e$-folds $N$ during the subsequent inflationary stage for any given choice of parameters $(p,q,m)$. 
Once $N$ and $(p,q,m)$ are specified, the tensor spectral index $n_T$ can be obtained from~\eqref{eq:nT value}, and the tensor amplitude $A_T$ can be computed using (see Ref.~\cite{Tahara:2020fmn} for the derivation)
\begin{align}
    \mathcal{P}_{\rm GW} 
    &= 2 \cdot \frac{k^3}{2\pi^2} \, |\tilde h_A|^2
     = \frac{2^{2\nu} \Gamma(\nu)^2}{\pi^3 M_{\mathrm{Pl}}^2} 
       \left( \frac{2 H_c}{2\nu-1} \right)^{2\nu-1} 
       k^{3-2\nu}  \notag \\
    &\equiv A_T \left( \frac{k}{k_{*}} \right)^{n_T}\;,
    \qquad 
    H_c \equiv H a^{-1/p}=p\cdot a_0^{-1/p}, \quad \nu=\frac{3p+1}{2(p+1)} \;.
\end{align}

\begin{figure}[h!]
    \centering
    \includegraphics[width=0.73\textwidth]{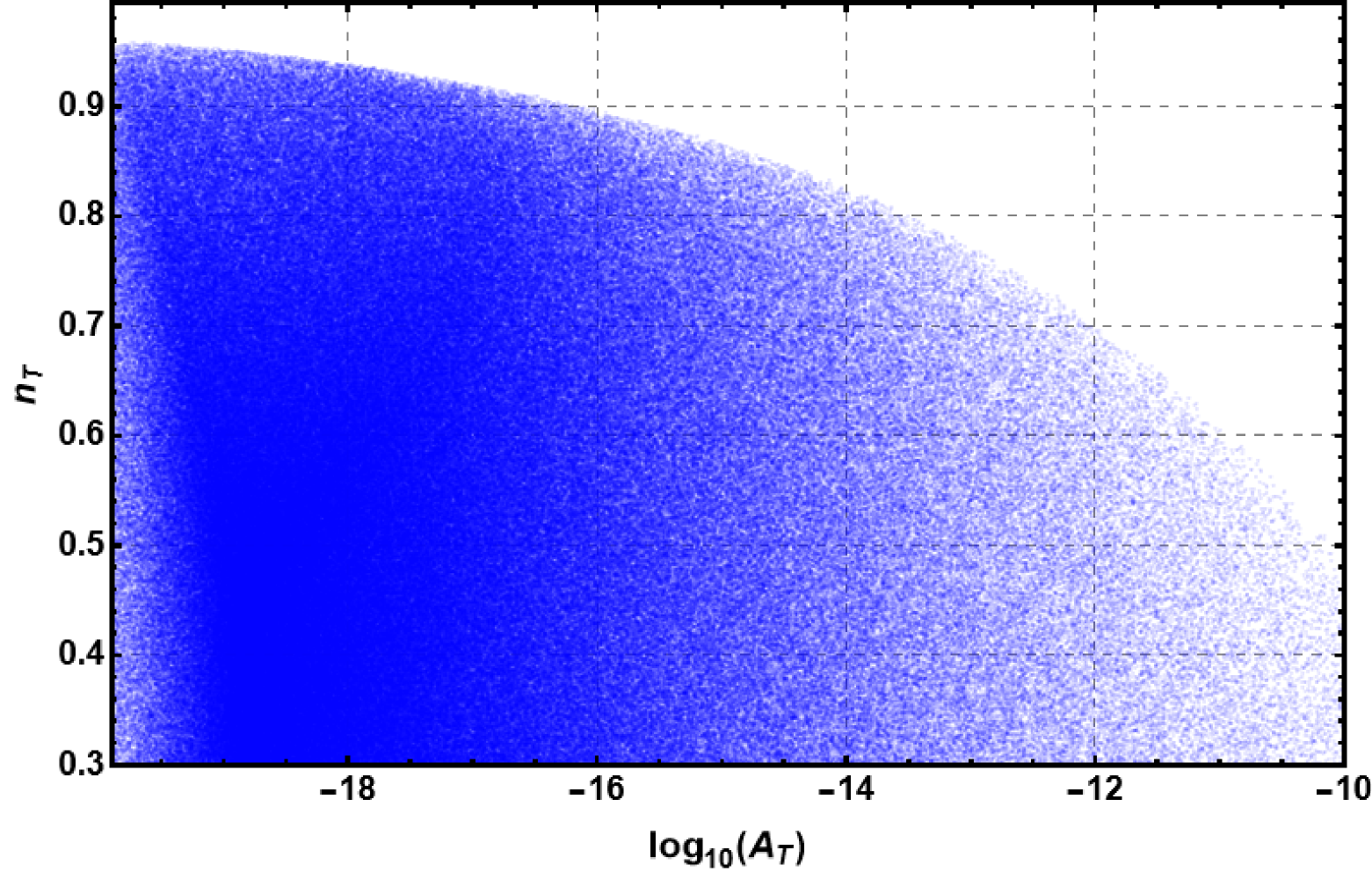}
    \caption{The blue region corresponds to the allowed values of the tensor spectral index $n_T$ and the tensor amplitude $A_T$. }
    \label{fig:nT and AT}
\end{figure}

We choose the pivot frequency 
\[
f_{*} \equiv \frac{k_{*}}{2\pi} = 10^{-8}\,\mathrm{Hz},
\] 
and generate $2\times 10^8$ sampling points with a uniform distribution in the parameter space 
\[
[p, q, \log_{10}(m/M_{\mathrm{Pl}})].
\] 
The sampled points are required to satisfy the following conditions: stability, absence of superluminal propagation, perturbative unitarity bounds~\eqref{eq:EstrEclass} and~\eqref{eq:ratio for cutoff}, and a number of $e$-folds during the subsequent inflationary stage in the range 
\[
N \in [1,60].
\] 
The resulting allowed region in the $(n_T, A_T)$ plane is shown in Fig.~\ref{fig:nT and AT}. We see that, after imposing perturbative unitarity constraints, typical values of $A_T$ lie in the range 
\[
A_T \sim [10^{-20},\,10^{-10}],
\] 
while typical values of $A_T$ from Ref.~\cite{Tahara:2020fmn} are in the range 
\[
A_T \sim [10^{-6},\,10^{-3}].
\] 
Hence, the perturbative unitarity bounds place strong constraints on the model parameters and reduce the allowed magnitude of $A_T$, which can be generated by the NEC-violating model of Ref.~\cite{Tahara:2020fmn}, by roughly four orders of magnitude.
We note that our estimation of the perturbative unitarity bounds is based on the requirement of perturbative unitarity in the high-energy limit. Equivalently, the similar bounds can be obtained by demanding that the cutoff derived from the Goldstone boson action (associated with the broken four-dimensional diffeomorphisms) in the decoupling limit lies well above the Hubble scale. 

Therefore, one may argue that the bounds derived here could be overly restrictive. It would thus be worthwhile to explore alternative ways of imposing unitarity in early-universe cosmology and to compare them with the bounds obtained in this work. For a discussion of possible alternative approaches, see Section~\ref{sec: Conclusion}.

\section{Conclusion}
\label{sec: Conclusion}

In this work, we revisited the cosmological scenario introduced in Ref.~\cite{Tahara:2020fmn}, which features an early phase of stable violation of the null energy condition (NEC) followed by a conventional inflationary era. This class of models is phenomenologically attractive: it evades Big Bang nucleosynthesis (BBN) constraints, allows a stable realization of NEC violation free of ghost and gradient instabilities, smoothly connects to a standard reheating stage, and generates a blue-tilted spectrum of primordial gravitational waves.

Despite these appealing features, the underlying theoretical framework raises important consistency issues. The model is formulated within Horndeski gravity, whose action generically contains higher-derivative operators. As a consequence, the effective field theory (EFT) describing cosmological perturbations is characterized by a background-dependent strong-coupling (dynamical cutoff) scale that lies parametrically below the Planck mass. This observation motivates a careful assessment of perturbative unitarity and the regime of validity of the theory.

To this end, we analyzed the quadratic and cubic actions for cosmological perturbations in the high-energy regime, defined by characteristic scattering times that are much shorter than the Hubble time $H^{-1}$, as well as shorter than the timescales associated with $H/\dot H$ and with the evolution of interaction vertices in the cubic Lagrangian. We refer to this regime as the \emph{high-energy limit}. In this limit, the theory—unlike conventional Lorentz-invariant quantum field theories—contains operators that explicitly violate Lorentz symmetry; however, it is still defined on a quasi-Minkowski background. Consequently, an $S$-matrix description is approximately well-defined in the high-energy limit, allowing for a systematic study of partial-wave amplitudes and perturbative unitarity bounds. We employed both power-counting arguments based on dimensional analysis and rigorous bounds derived from the optical theorem. 

Our main conclusion is the following: requiring perturbative control of cosmological perturbations—namely, that all relevant energy scales remain parametrically below the dynamical cutoff—severely restricts the amplitude of primordial gravitational waves that can be produced during the NEC-violating phase. In particular, within this framework it is highly challenging, without entering a strong-coupling regime, to generate a nanohertz gravitational-wave signal of sufficiently large amplitude. In fact, the typical amplitude is found to be roughly four orders of magnitude lower than one might naively expect.

Moreover, we do not expect the strong-coupling scale identified in this work to be a gauge artifact. Gauge-related subtleties typically disappear in the decoupling (high-energy) limit, \( H \to 0 \). A potential caveat is that the decoupling limit itself may be ill-defined, which can occur when inverse powers of the Hubble parameter appear in the interaction action; see Ref.~\cite{deRham:2017aoj} for a discussion. However, this issue does not arise in the present analysis: all terms considered here remain well defined in the decoupling limit \( H \to 0 \). In addition, our method reproduces the same energy scaling of the cutoff as analyses based on the decoupling limit of the Goldstone boson action (corresponding to the broken four-dimensional diffeomorphism invariance), providing an additional consistency check of our approach. See the discussion at the end of Section~\ref{subsec: Matrix Element for the Leading-Order Action}.

Importantly, our conclusions do not rely on the detailed realization of the transition between the NEC-violating phase and the subsequent inflationary stage. The analysis therefore applies to a broader class of cosmological models. Furthermore, the violation of perturbative unitarity does not necessarily imply a breakdown of the underlying theory itself; rather, it may simply signal the inapplicability of the perturbative description, as discussed, for example, in Refs.~\cite{Aydemir:2012nz, Karananas:2022byw}. An important direction for future work is thus to employ non-perturbative methods to understand the dynamics beyond the perturbative regime.

In this article, we focused on a particular realization of an NEC-violating phase. A natural extension of this work is to analyze more general setups within Horndeski gravity and to investigate the bounds that unitarity can, in principle, impose on the amplitude of primordial gravitational waves. A promising approach is to employ the effective field theory (EFT) description of Horndeski models~\cite{Kase:2014cwa, Kennedy:2017sof}. We also expect that a similar analysis can be applied to Generalized Proca cosmologies by adapting the EFT framework for vector--tensor theories~\cite{Aoki:2021wew} and subsequently studying perturbative unitarity bounds.

Finally, our analysis relies on the high-energy limit, in which scattering times are much shorter than the Hubble time. This approach breaks down when scattering times become comparable to $H^{-1}$, in which case an $S$-matrix description is no longer well-defined. More generally, defining scattering amplitudes in spacetimes that are not asymptotically flat is intrinsically challenging (see, e.g., Refs.~\cite{Mack:2009mi, Penedones:2010ue, Marolf:2012kh, Melville:2023kgd, Donath:2024utn, Melville:2024ove} for attempts in de Sitter and anti--de Sitter spaces). One may therefore argue that the bounds derived here are overly restrictive. An alternative approach is to consider constraints based on momentum-space entanglement; see Ref.~\cite{DuasoPueyo:2024usw} for related work. Such bounds are expected to be weaker than those obtained in the present analysis, although a detailed comparison remains to be carried out. It would also be interesting to investigate whether the cosmological optical theorem~\cite{Goodhew:2020hob} can be employed to derive constraints on the model considered here. A systematic comparison of the bounds obtained from these different methods is left for future work.

\section*{ACKNOWLEDGMENTS}

We are grateful to Mohammad Ali Gorji, Tsutomu Kobayashi, Yusuke Yamada, Shengfeng Yan and Mian Zhu for useful discussion and comments. P.P. is supported by IBS under the project code, IBS-R018-D3. J.W. is supported in part by JSPS KAKENHI No.~24K00624. J.W. is also supported by Kavli IPMU, which was established by the World Premier International Research Center
Initiative (WPI), MEXT, Japan.

\newpage
\appendix

\section{The Cubic Actions for All Sectors}
\label{app: The cubic action for all sectors}
\numberwithin{equation}{section}

\subsection{Scalar Sector}

The quadratic action for scalar perturbations is given by
\begin{equation}
\mathcal{S}_{\mathcal{R}\mathcal{R}}^{(2)} = \int \mathrm{d} t \mathrm{~d}^3 x N a^3\left[\mathcal{G}_S \frac{\dot{\mathcal{R}}^2}{N^2}-\frac{\mathcal{F}_S}{a^2} \mathcal{R}_{, i} \mathcal{R}_{, i}\right],
\end{equation}
where $\mathcal{F}_S$ and $\mathcal{G}_S$ are gradient term coefficient and kinetic term coefficient, respectively. 
Using the freedom of time reparameterization, one can, without loss of generality, set the lapse function to unity, that is, $N = 1$. After this choice, cosmic time $t_c = \int N(t) \, \mathrm{d}t$ coincides with the coordinate time $t$.

The general expressions for the coefficients $\mathcal{F}_S$ and $\mathcal{G}_S$ are given by
\begin{equation}\label{def:FSGS}
\mathcal{F}_S = \frac{1}{a} \frac{\mathrm{d}}{\mathrm{d}t} \left( \frac{a}{\Theta} \mathcal{G}_T^2 \right) - \mathcal{F}_T, \quad 
\mathcal{G}_S = \frac{\Sigma}{\Theta^2} \mathcal{G}_T^2 + 3 \mathcal{G}_T,
\end{equation}
and
\begin{equation}
\label{eq:FTFS}
\mathcal{F}_T = 2G_4 = M_{\mathrm{Pl}}^2, \quad 
\mathcal{G}_T = 2G_4 = M_{\mathrm{Pl}}^2,
\end{equation}
where
\begin{equation}
\begin{aligned}\label{eq:A4}
\Sigma =\ & X G_{2X}  + 12 H \dot{\phi} X G_{3X} - 2 X G_{3\phi} - 2 X^2 G_{3\phi X} -6 H^2 G_4\\
=\, & X \beta+12 H \dot{\phi} X \frac{\alpha M_{\mathrm{Pl}}}{U(\phi)}+4 X^2 \frac{\alpha M_{\mathrm{Pl}} U^{\prime}(\phi)}{U^2(\phi)}-3 H^2 M_{\mathrm{Pl}}^2 \\
=\, & \frac{2 \left(9 p^2+6 p+1\right) p q-9 p-3}{(3 p+1) q} \frac{M_{\mathrm{Pl}}^2}{t^2}, \\
\Sigma_X =\ & \beta + 12 H \dot{\phi} \frac{\alpha M_{\mathrm{Pl}}}{U(\phi)} + 8 X \frac{\alpha M_{\mathrm{Pl}} U'(\phi)}{U^2(\phi)} \\
=\, & \frac{6 p (-((-3 p-2) (3 p+1) q)-3)-14}{(3 p+1) q} , \\
\Theta =\ & -\dot{\phi} X G_{3X} + 2H G_4 \\
=\, & -\frac{1}{(3p + 1)q} \frac{M_{\mathrm{Pl}}^2}{t}.
\end{aligned}
\end{equation}
For the current model, we have $p>1$ and $q>0$. With this choice of parameter ranges, all of the expressions above are well-defined and never divergent.  
Substituting Eq.\eqref{eq:A4}, we arrive at:
\begin{equation}
\begin{aligned}\label{eq:FSGS}
\mathcal{F}_S &= M_{\mathrm{Pl}}^2 \left[ (p - 1)(3p + 1)q - 1 \right], \\
\mathcal{G}_S &= M_{\mathrm{Pl}}^2 \left[ 3+(1+3 p)^2 q(-3+2 p(1+3 p) q) \right].
\end{aligned}
\end{equation}
Now let us turn to the cubic action for scalar perturbations:
\begin{equation}
\begin{aligned}
\mathcal{S}_{\mathcal{R} \mathcal{R} \mathcal{R}}^{(3)}= & \int a^4 \mathrm{d} \eta \mathrm{d}^3 x\left\{\Lambda_1 \dot{\mathcal{R}}^3 +\Lambda_2 \dot{\mathcal{R}}^2 \mathcal{R}+\Lambda_3 \dot{\mathcal{R}}^2 \partial^2 \mathcal{R}+\Lambda_4 \dot{\mathcal{R}} \mathcal{R} \partial^2 \mathcal{R}\right. \\
& +\Lambda_5 \dot{\mathcal{R}}\left(\partial_i \mathcal{R}\right)^2+\Lambda_6 \mathcal{R}\left(\partial_i \mathcal{R}\right)^2+\Lambda_7 \dot{\mathcal{R}}\left(\partial^2 \mathcal{R}\right)^2+\Lambda_8 \mathcal{R}\left(\partial^2 \mathcal{R}\right)^2+\Lambda_9 \partial^2 \mathcal{R}\left(\partial_i \mathcal{R}\right)^2 \\
& +\Lambda_{10} \dot{\mathcal{R}}\left(\partial_i \partial_j \mathcal{R}\right)^2+\Lambda_{11} \mathcal{R}\left(\partial_i \partial_j \mathcal{R}\right)^2+\Lambda_{12} \dot{\mathcal{R}} \partial_i \mathcal{R} \partial_i \psi+\Lambda_{13} \partial^2 \mathcal{R} \partial_i \mathcal{R} \partial_i \psi+\Lambda_{14} \dot{\mathcal{R}}\left(\partial_i \partial_j \psi\right)^2 \\
& \left.+\Lambda_{15} \mathcal{R}\left(\partial_i \partial_j \psi\right)^2+\Lambda_{16} \dot{\mathcal{R}} \partial_i \partial_j \mathcal{R} \partial_i \partial_j \psi+\Lambda_{17} \mathcal{R} \partial_i \partial_j \mathcal{R} \partial_i \partial_j \psi\right\},
\end{aligned}
\end{equation}
where $\partial^2=\partial_i \partial_i$,
\begin{equation}
\psi=\partial^{-2} \dot{\mathcal{R}}.
\end{equation}
The expressions for all the coefficients \(\Lambda_{1} \sim \Lambda_{17}\) are given below:
\begin{equation}
\begin{aligned}
\Lambda_1 = & -\frac{\mathcal{G}_T^3}{3 \Theta^3}\left(\Sigma+2 X \Sigma_X+H \Xi\right)+\frac{\mathcal{G}_T^2 \Xi}{\Theta^2}-\frac{\mathcal{G}_T \mathcal{G}_S \Xi}{3 \Theta^2}+\frac{\Gamma \mathcal{G}_S^2}{2 \Theta \mathcal{G}_T} -\frac{2 \Gamma \mathcal{G}_S}{\Theta}+\frac{3 \Gamma \mathcal{G}_T}{\Theta}, \\
\Lambda_2 =&  \frac{3 \mathcal{G}_T^2 \Sigma}{\Theta^2}+9 \mathcal{G}_T-\frac{3 \mathcal{G}_S^2}{2 \mathcal{G}_T}, \;\;
\Lambda_3 =  \frac{\mathcal{G}_T^3 \Xi}{3 a^2 \Theta^3}-\frac{\mathcal{G}_T \mathcal{G}_S \Gamma}{a^2 \Theta^2}+\frac{2 \Gamma \mathcal{G}_T^2}{a^2 \Theta^2},\\
\Lambda_4   =& \frac{3 \mathcal{G}_T \mathcal{G}_S}{a^2 \Theta}-\frac{2 \mathcal{G}_T^2}{a^2 \Theta}, \;\;
\Lambda_5 =-\frac{\mathcal{G}_T^2}{a^2 \Theta}+\frac{2 \mathcal{G}_T \mathcal{G}_S}{a^2 \Theta},\;\;
\Lambda_6  =\frac{\mathcal{F}_T}{a^2} ,\\
\Lambda_7 &=\frac{\Gamma \mathcal{G}_T^3}{2 a^4 \Theta^3} ,\;\;
\Lambda_8 =-\frac{3 \mathcal{G}_T^3}{2 a^4 \Theta^2}, \;\;
\Lambda_9 =-\frac{2 \mathcal{G}_T^3}{a^4 \Theta^2},\\
\Lambda_{10} & =-\frac{\Gamma \mathcal{G}_T^3}{2 a^4 \Theta^3}, \;\;
\Lambda_{11} =\frac{3 \mathcal{G}_T^3}{2 a^4 \Theta^2},\;\;
\Lambda_{12}  =-\frac{2 \mathcal{G}_S^2}{\mathcal{G}_T}, \\
\Lambda_{13} &=\frac{2 \mathcal{G}_T \mathcal{G}_S}{a^2 \Theta},\;\;
\Lambda_{14}  =-\frac{\Gamma \mathcal{G}_S^2}{2 \Theta \mathcal{G}_T} ,\;\;
\Lambda_{15} =\frac{3 \mathcal{G}_S^2}{2 \mathcal{G}_T},\\
\Lambda_{16} & =\frac{\mathcal{G}_T \mathcal{G}_S \Gamma}{a^2 \Theta^2}, \;\;
\Lambda_{17} =-\frac{3 \mathcal{G}_T \mathcal{G}_S}{a^2 \Theta} .
\end{aligned}
\end{equation}
where
\begin{equation}
\begin{aligned}\label{eq:KGB}
\Xi & =12 \frac{\dot{\phi}}{N} X G_{3 X} -12 H G_4 = - \frac{6(-2+p(1+3 p) q)}{(1+3 p) q} \frac{M_{\mathrm{Pl}}^2}{ t}\\
\Gamma & =\mathcal{G}_T=\mathcal{F}_T=2 G_4 =M_{\mathrm{Pl}}^2, \\
\mu & =0.
\end{aligned}
\end{equation}
The model in Eq.\eqref{eq: Tahara model covariant Lagrangian} belongs to a subclass of Horndeski gravity, commonly referred to as Kinetic Gravity Braiding (KGB) \cite{Deffayet:2010qz}. For this subclass, the coefficients $\Lambda_i$ in the cubic Lagrangian for scalar perturbations satisfy the following relations:
\begin{equation}\label{eq:sameLambda}
\Lambda_{11}=-\Lambda_{8}, \quad \Lambda_{10}=-\Lambda_{7}.
\end{equation}
We have used this in deriving Eq.\eqref{eq: L} and Eq.\eqref{eq: SL}.
Substituting Eq.\eqref{def:FSGS}, \eqref{eq:FTFS}, \eqref{eq:A4} and \eqref{eq:KGB}, the expression for all 17 terms can be organized as
\begin{equation}
\begin{aligned}
&\Lambda_1 /M_{\mathrm{Pl}}^2  =\frac{\lambda_1}{H},\;\;
\Lambda_2 /M_{\mathrm{Pl}}^2 =\lambda_2,\;\;
\Lambda_3 /M_{\mathrm{Pl}}^2  = \frac{\lambda_3}{a^2 H^2},\;\; \Lambda_4 /M_{\mathrm{Pl}}^2 = \frac{\lambda_4 }{a^2 H}, \;\;
\Lambda_5 /M_{\mathrm{Pl}}^2  = \frac{\lambda_5 }{a^2 H} ,\;\; \\
&
\Lambda_6 /M_{\mathrm{Pl}}^2 = \frac{\lambda_6 }{a^2},\;\; \Lambda_7 /M_{\mathrm{Pl}}^2 = \frac{\lambda_7 }{a^4 H^3},\;\;
\Lambda_8 /M_{\mathrm{Pl}}^2 = \frac{\lambda_8 }{a^4 H^2} ,\;\;
\Lambda_9 /M_{\mathrm{Pl}}^2  = \frac{\lambda_9 }{a^4 H^2},\;\;\Lambda_{12} /M_{\mathrm{Pl}}^2  =\lambda_{12}, \\
&
\;\;\Lambda_{13} /M_{\mathrm{Pl}}^2 =  \frac{\lambda_{13}}{a^2 H} ,\;\;
\Lambda_{14} /M_{\mathrm{Pl}}^2 =\frac{\lambda_{14} }{H},\;\;\Lambda_{15} /M_{\mathrm{Pl}}^2 =\lambda_{15}, \;\;\Lambda_{16} /M_{\mathrm{Pl}}^2 = \frac{\lambda_{16} }{a^2 H^2},\;\;
\Lambda_{17} /M_{\mathrm{Pl}}^2  =  \frac{\lambda_{17}}{a^2 H},
\end{aligned}
\end{equation}
where $\lambda_i$ are give as:
\begin{equation}
\begin{aligned}
&\lambda_1 = \frac{p(1+3 p)q}{6}\left[9+(1+3p)q \cdot \right.\\ 
&\left.\left(p \left(12 p (3 p+1)^5 q^3-12 (11 p+3) (3 p+1)^3 q^2+(9 p (157 p+129)+311) q-192\right)+27 q-56\right)\right],\\
&\lambda_2 = \frac{3}{2} \left\{ 2 (3 p+1)^2 q (2 p (3 p+1) q-3)-\left((3 p+1)^2 q (2 p (3 p+1) q-3)+3\right)^2+6 \right\}, \\
&\lambda_3 = - p^2 (3 p+1)^2 q^2\left(q\left(p\left(2(3 p+1)^3 q-33 p-20\right)-3\right)+5\right),\\
&\lambda_4 = p(3 p+1) q\left(3(3 p+1)^2 q(2 p(3 p+1) q-3)+7\right),\\
&\lambda_5 =p (3 p+1) q\left(2(3 p+1)^2 q(2 p(3 p+1) q-3)+5\right),\\
&\lambda_6 =1, \quad 
\lambda_7 = \frac{1}{2} p^3 (3 p q+q)^3,\quad \lambda_8 =- \frac{3}{2} p^2 (3 p+1)^2 q^2,\quad \lambda_9 =-2 p^2 (3 p+1)^2 q^2,\\
&\lambda_{12} =-2 \left((3 p+1)^2 q(2 p(3 p+1) q-3)+3\right)^2,\\
&\lambda_{13} = 2 p(3 p+1) q\left((3 p+1)^2 q(2 p(3 p+1) q-3)+3\right),\\
&\lambda_{14} = -\frac{1}{2} p(3 p+1) q\left((3 p+1)^2 q(2 p(3 p+1) q-3)+3\right)^2,\\
&\lambda_{15} = \frac{3}{2} \left((3 p+1)^2 q(2 p(3 p+1) q-3)+3\right)^2 ,\\
&\lambda_{16} = p^2(3 p+1)^2 q^2\left((3 p+1)^2 q(2 p(3 p+1) q-3)+3\right) ,\\
&\lambda_{17} = - 3 p(3 p+1) q\left((3 p+1)^2 q(2 p(3 p+1) q-3)+3\right).
\end{aligned}
\end{equation}


\subsection{Mixed and Tensor Sectors}

The quadratic action for transverse–traceless (spin-2) tensor perturbations reads
\begin{equation}
\mathcal{S}_{h h}^{(2)}=\int \mathrm{~d} t \mathrm{~d}^3 x N \frac{a^3}{8}\left[\mathcal{G}_T \frac{\dot{h}_{i j}^2}{N^2}-\frac{\mathcal{F}_T}{a^2} h_{i j, k} h_{i j, k}\right].
\end{equation}
In the following, we present the cubic actions for perturbations in the mixed and tensor sectors. The action for the tensor–tensor–scalar sector is given by:  
\begin{equation}
\begin{aligned}
\mathcal{L}_{\mathcal{R} h h}^{(3)} =& a^3\left[d_1 \mathcal{R} \frac{\dot{h}_{i j}^2}{N^2}+\frac{d_2}{a^2} \mathcal{R} h_{i j, k} h_{i j, k}+d_3 \psi_{, k} \frac{\dot{h}_{i j}}{N} h_{i j, k} \right. \\
& \left.+d_4 \frac{\dot{\mathcal{R}}}{N} \frac{\dot{h}_{i j}^2}{N^2} +\frac{d_5}{a^2} \partial^2 \mathcal{R} \frac{\dot{h}_{i j}^2}{N^2} +d_6 \psi_{, i j} \frac{\dot{h}_{i k}}{N} \frac{\dot{h}_{j k}}{N}+\frac{d_7}{a^2} \mathcal{R}_{, i j} \frac{\dot{h}_{i k}}{N} \frac{\dot{h}_{j k}}{N}
\right] .
\end{aligned}
\end{equation}
For KGB subclass, we have \eqref{eq:KGB}.
Additionally, we have $c_T^2 \equiv \mathcal{F}_T / \mathcal{G}_T \mathcal{}= 1$, and
\begin{equation}
d_4=d_5=d_6=d_7=0.
\end{equation}
Thus we arrive at
\begin{equation}
\begin{aligned}
d_1 & =\frac{3 \mathcal{G}_T}{8}\left[1-\frac{H \mathcal{G}_T^2}{\Theta \mathcal{F}_T}+\frac{\mathcal{G}_T}{3} \frac{\mathrm{d}}{N \mathrm{d} t}\left(\frac{\mathcal{G}_T}{\Theta \mathcal{F}_T}\right)\right] = \widetilde{d}_1 M_{\mathrm{Pl}}^2,\\
d_2 & =\frac{\mathcal{F}_S}{8} = \widetilde{d}_2 M_{\mathrm{Pl}}^2,\\
d_3 & =-\frac{\mathcal{G}_S}{4} = \widetilde{d}_3 M_{\mathrm{Pl}}^2,
\end{aligned}
\end{equation}
where
\begin{equation}
\begin{aligned}
\widetilde{d}_1 \equiv d_1 / M_{\mathrm{Pl}}^2& =-\frac{1}{8} \left(-3+(1+3p)^2 q\right),\\
\widetilde{d}_2 \equiv d_2 / M_{\mathrm{Pl}}^2 & =\frac{1}{8} [(p-1)(3 p+1) q-1], \\
\widetilde{d}_3 \equiv d_3 / M_{\mathrm{Pl}}^2 & =-\frac{1}{4} \left[3+(1+3 p)^2 q(-3+2 p(1+3 p) q)\right] .
\end{aligned}
\end{equation}
Next, let us turn to the tensor-scalar-scalar sector
\begin{equation}
\begin{aligned}
\mathcal{L}_{\mathcal{R} \mathcal{R} h}^{(3)}= & a^3\left[\frac{c_1}{a^2} h_{i j} \mathcal{R}_{, i} \mathcal{R}_{, j}+\frac{c_2}{a^2} \frac{\dot{h}_{i j}}{N} \mathcal{R}_{, i} \mathcal{R}_{, j}+c_3 \frac{\dot{h}_{i j}}{N} \mathcal{R}_{, i} \psi_{, j}\right. \\
& \left.+\frac{c_4}{a^2} \partial^2 h_{i j} \mathcal{R}_{, i} \psi_{, j} +\frac{c_5}{a^4} \partial^2 h_{i j} \mathcal{R}_{, i} \mathcal{R}_{, j}+c_6 \partial^2 h_{i j} \psi_{, i} \psi_{, j}\right] .
\end{aligned}
\end{equation}
For our model, we have
\begin{equation}
\begin{aligned}
c_1= & \mathcal{F}_S =\widetilde{c}_1 M_\mathrm{Pl}^2, \\
c_2= & \frac{\Gamma}{4 \Theta}\left(\mathcal{F}_S-\mathcal{F}_T\right)+\frac{\mathcal{G}_T^2}{\Theta}\left[-\frac{1}{2}+\frac{3 H \Gamma}{4 \Theta}-\frac{1}{4} \frac{\mathrm{d}}{N\mathrm{d} t}\left(\frac{\Gamma}{\Theta}\right)\right] =\frac{\widetilde{c}_2}{H} M_\mathrm{Pl}^2,\\
c_3= & \mathcal{G}_S\left[\frac{3}{2}+\frac{\mathrm{d}}{N \mathrm{d} t}\left(\frac{\Gamma}{2 \Theta}\right)-3 H \frac{\Gamma}{2 \Theta}\right] =\widetilde{c}_3 M_\mathrm{Pl}^2,\\
c_4= & 0, \quad 
c_5= 0, \quad 
c_6= \frac{\mathcal{G}_S^2}{4 \mathcal{G}_T} =\widetilde{c}_6 M_\mathrm{Pl}^2.
\end{aligned}
\end{equation}
where
\begin{equation}
\begin{aligned}
\widetilde{c}_1\equiv c_1/M_\mathrm{Pl}^2 = & (p-1)(3 p+1) q-1, \\
\widetilde{c}_2 \equiv c_2 H/M_\mathrm{Pl}^2 
= & p(3 p+1) q (p (3 p+1) q-1), \\
\widetilde{c}_3\equiv c_3/M_\mathrm{Pl}^2 
= & -\frac{1}{2} \left((3 p+1)^2 q-3\right) \left((3 p+1)^2 q (2 p (3 p+1) q-3)+3\right),\\
\widetilde{c}_6\equiv c_6/M_\mathrm{Pl}^2 = &\frac{1}{4} \left[3+(1+3 p)^2 q(-3+2 p(1+3 p) q)\right]^2.
\end{aligned}
\end{equation}
Finally, the cubic action for pure tensor perturbations is expressed in the following form: 
\begin{equation}
\mathcal{L}_{h h h}^{(3)}=a^3 \frac{\mathcal{F}_T}{4a^2 }\left(h_{i k} h_{j l}-\frac{1}{2} h_{i j} h_{k l}\right) h_{i j, k l} .
\end{equation}

\section{Dimensional Analysis}
\label{app: Dimensional analysis}
\numberwithin{equation}{section}

\subsection{Scalar Sector}

Firstly, we perform a redefinition of the variables and a canonical normalization of the field:
\begin{equation}
\begin{aligned}\label{def:norcanoS}
&\mathrm{d}t=a\mathrm{d}\eta, \quad \partial_t=a^{-1} \partial_{\eta}, \quad H=\frac{\dot{{a}}}{a}=\frac{a'}{a^2},\\
\mathcal{R} =& a^{-1} \left(\sqrt{2\mathcal{G}_S}\right)^{-1} \mathcal{R}_c , \quad 
\dot{\mathcal{R}} = a^{-1} \left(\sqrt{2\mathcal{G}_S}\right)^{-1} \left( a^{-1} \mathcal{R}_c^{\prime}- H \mathcal{R}_c \right).
\end{aligned}
\end{equation}
Additionally, for simplicity, we introduce the following auxiliary variables: 
\begin{equation}
\psi_a \equiv \partial^{-2}\mathcal{R}_c^{\prime}\;,\quad 
\phi_a \equiv \partial^{-2}\mathcal{R}_c\;,
\end{equation}
so that we can express $\psi$ as
\begin{equation}
\psi= a^{-1} \left(\sqrt{2\mathcal{G}_S}\right)^{-1} \left( a^{-1} \psi_a- H \phi_a \right).
\end{equation}
The quadratic action for scalar perturbations after canonical normalization reads as Eq.\eqref{eq:canoscalar}.
Having Eq.\eqref{def:norcanoS}, we can rewrite the cubic action in terms of the canonically normalized field:
\begin{equation}
\begin{aligned}\label{eq:scalar3c}
\mathcal{S}_{\mathcal{R} \mathcal{R} \mathcal{R}}^{(3)}= & \left(\sqrt{2\mathcal{G}_S}\right)^{-3} \int \mathrm{d} \eta \mathrm{d}^3 x\left\{\Lambda_1  \left(a^{-2} \left(\mathcal{R}_c^{\prime}\right)^3 -3H a^{-1} \left(\mathcal{R}_c^{\prime}\right)^2 \mathcal{R}_c + 3H^2 ~ \mathcal{R}_c^{\prime} \mathcal{R}_c^2 -H^3 a ~\mathcal{R}_c^3\right) \right. \\
& +\Lambda_2 \left(a^{-1} \left(\mathcal{R}_c^{\prime}\right)^2 \mathcal{R}_c-2 H ~\mathcal{R}_c^{\prime}\mathcal{R}_c^2 + H^2  a ~\mathcal{R}_c^3\right) \\
&+\Lambda_3 \left(a^{-1} \left(\mathcal{R}_c^{\prime}\right)^2 \partial^2 \mathcal{R}_c-2 H ~\mathcal{R}_c^{\prime} \mathcal{R}_c \partial^2 \mathcal{R}_c+ H^2 a~\mathcal{R}_c^2 \partial^2 \mathcal{R}_c\right) \\
&+\Lambda_4 \left(\mathcal{R}_c^{\prime} \mathcal{R}_c \partial^2 \mathcal{R}_c -H a~\mathcal{R}_c^2 \partial^2 \mathcal{R}_c\right)  +\Lambda_5 \left(\mathcal{R}_c^{\prime} \left(\partial_i \mathcal{R}_c\right)^2-H a~\mathcal{R}_c \left(\partial_i \mathcal{R}_c\right)^2 \right)  +\Lambda_6 ~a~ \mathcal{R}_c\left(\partial_i \mathcal{R}_c\right)^2\\
& +\Lambda_7 \left( \mathcal{R}_c^{\prime}\left(\partial^2 \mathcal{R}_c\right)^2-H a~\mathcal{R}_c\left(\partial^2 \mathcal{R}_c\right)^2\right)  +\Lambda_8 ~a~ \mathcal{R}_c\left(\partial^2 \mathcal{R}_c\right)^2+\Lambda_9 ~a~ \partial^2 \mathcal{R}_c\left(\partial_i \mathcal{R}_c\right)^2 \\
& +\Lambda_{10} \left(\mathcal{R}_c^{\prime}\left(\partial_i \partial_j \mathcal{R}_c\right)^2-H a~\mathcal{R}_c\left(\partial_i \partial_j \mathcal{R}_c\right)^2\right)  +\Lambda_{11} ~a~\mathcal{R}_c\left(\partial_i \partial_j \mathcal{R}_c\right)^2\\
& +\Lambda_{12} \left(a^{-1}\mathcal{R}_c^{\prime} \partial_i \mathcal{R}_c \partial_i \psi_a-H ~\mathcal{R}_c \partial_i \mathcal{R}_c \partial_i \psi_a -H ~\mathcal{R}_c^{\prime} \partial_i \mathcal{R}_c \partial_i \partial^{-2} \mathcal{R}_c+H^2 a~\mathcal{R}_c \partial_i \mathcal{R}_c \partial_i \partial^{-2} \mathcal{R}_c\right) 
\\
& +\Lambda_{13} \left(\partial^2 \mathcal{R}_c \partial_i \mathcal{R}_c \partial_i \psi_a-H a~ \partial^2 \mathcal{R}_c \partial_i \mathcal{R}_c \partial_i \partial^{-2} \mathcal{R}_c\right)\\
& +\Lambda_{14} \left(a^{-2} \mathcal{R}_c^{\prime}\left(\partial_i \partial_j \psi_a\right)^2-2H a^{-1} \mathcal{R}_c^{\prime}\left(\partial_i \partial_j \psi_a\right) \left(\partial_i \partial_j \partial^{-2} \mathcal{R}_c\right) +H^2 ~\mathcal{R}_c^{\prime}\left(\partial_i \partial_j \partial^{-2} \mathcal{R}_c\right)^2 \right) \\
& +\Lambda_{14} \left(-H a^{-1} \mathcal{R}_c\left(\partial_i \partial_j \psi_a\right)^2+2H^2 \mathcal{R}_c\left(\partial_i \partial_j \psi_a\right) \left(\partial_i \partial_j \partial^{-2} \mathcal{R}_c\right) -H^3 a \mathcal{R}_c \left(\partial_i \partial_j \partial^{-2} \mathcal{R}_c\right)^2 \right) \\
& +\Lambda_{15} \left(a^{-1} \mathcal{R}_c \left(\partial_i \partial_j \psi_a\right)^2-2H ~\mathcal{R}_c \left(\partial_i \partial_j \psi_a\right) \left(\partial_i \partial_j \partial^{-2} \mathcal{R}_c\right) +H^2 a~ \mathcal{R}_c \left(\partial_i \partial_j \partial^{-2} \mathcal{R}_c\right)^2 \right)\\
& +\Lambda_{16}\left( a^{-1}\mathcal{R}_c^{\prime}\left(\partial_i \partial_j \mathcal{R}_c\right)\partial_i \partial_j \psi_a- H~\mathcal{R}_c\left(\partial_i \partial_j \mathcal{R}_c\right)\partial_i \partial_j \psi_a \right) \\
& +\Lambda_{16}\left(- H ~\mathcal{R}'_c\left(\partial_i \partial_j \mathcal{R}_c\right)\partial_i \partial_j \partial^{-2} \mathcal{R}_c+H^2 a~\mathcal{R}_c \left(\partial_i \partial_j \mathcal{R}_c\right)\partial_i \partial_j \partial^{-2} \mathcal{R}_c \right) \\
& \left.+\Lambda_{17} \left(\mathcal{R}_c \left(\partial_i \partial_j \mathcal{R}_c\right)\left(\partial_i \partial_j \psi_a\right)-H a~\mathcal{R}_c \left(\partial_i \partial_j \mathcal{R}_c\right)\left(\partial_i \partial_j \partial^{-2} \mathcal{R}_c\right)\right) \right\}.
\end{aligned}
\end{equation}
Finally, let us perform a dimensional analysis and estimate the strong-coupling scale for each cubic interaction term in scalar sector. Additionally, we adopt the following approximation in our dimensional analysis: 
\begin{equation}
\begin{aligned}\label{eq:atoH}
a =&a_0(-t)^{-p}=a_0\left(\frac{p}{H}\right)^{-p}=a_0 p^{-p} H^p \sim a_0 H^p.
\end{aligned}
\end{equation}
We note that each interaction cubic term schematically can be rewritten in the following form:
\begin{align}\label{def:tildeLambda}
    &\widetilde{\Lambda}_{i} (\partial^n) \mathcal{R}_c^3\;,\;\;\widetilde{\Lambda}_{i} \sim \left(\sqrt{ \mathcal{G}_S}\right)^{-3}\Lambda_{i}  H^{x} a^{y} \sim M_{\mathrm{Pl}}^{-3} \Lambda_{i}  H^{x} a^{y}\;.
\end{align}
Here, $x$ and $y$ denote the powers of $H$ and $a$ in an arbitrary term of the cubic scalar action \eqref{eq:scalar3c}.

After that, using formula \eqref{eq: estimate strong-coupling scale} we estimate the strong-coupling energy scale for each term as follows.
\begin{equation*}
\begin{aligned}
   \boxed{\text{term }01(1)} \;  &\left(\mathcal{R}_c^{\prime}\right)^3: \; 
E_{\text{str}} \sim a_0 M_{\mathrm{Pl}}^{1/2} H^{p + 1/2}\;,\;\;
\boxed{\text{term }01(2)} \;  \left(\mathcal{R}_c^{\prime}\right)^2 \mathcal{R}_c: \; E_{\text{str}} \sim a_0 M_{\mathrm{Pl}} H^{p}\;,\;\;\\
\boxed{\text{term }01(3)} \;  &\mathcal{R}_c^{\prime} \mathcal{R}_c^2: \;  [\mathcal{R}_c^{\prime} \mathcal{R}_c^2] = 4\;,\;\;
\boxed{\text{term }01(4)} \;  \mathcal{R}_c^3: \; [\mathcal{R}_c^3 ]  = 3\;,
\end{aligned}
\end{equation*}

\begin{equation*}
\begin{aligned}
\boxed{\text{term }02(1)} \;  &\left(\mathcal{R}_c^{\prime}\right)^2 \mathcal{R}_c: E_{\text{str}} \sim a_0 M_{\mathrm{Pl}} H^{p}\;,\;\; 
\boxed{\text{term }02(2)} \;  \mathcal{R}_c^{\prime} \mathcal{R}_c^2: \;
[\mathcal{R}_c^{\prime} \mathcal{R}_c^2] = 4\;,\\
\boxed{\text{term }02(3)} \;  &\mathcal{R}_c^3: \; [\mathcal{R}_c^3]=3\;,\;\;
\boxed{\text{term }03(1)} \;  \left(\mathcal{R}_c^{\prime}\right)^2 \partial^2 \mathcal{R}_c: \; E_{\text{str}} \sim a_0 M_{\mathrm{Pl}}^{1/3} H^{2/3+p}\;,\\ 
\boxed{\text{term }03(2)} \; &\mathcal{R}_c^{\prime} \mathcal{R}_c \partial^2 \mathcal{R}_c: \; E_{\text{str}} \sim a_0 M_{\mathrm{Pl}}^{1/2} H^{p + 1/2} \;,\;\; 
\boxed{\text{term }03(3)} \; \mathcal{R}_c^2 \partial^2 \mathcal{R}_c: \; E_{\text{str}} \sim a_0 M_{\mathrm{Pl}} H^{p} \;,\\ 
\boxed{\text{term }04(1)} \;  &\mathcal{R}_c^{\prime} \mathcal{R}_c \partial^2 \mathcal{R}_c: \; E_{\text{str}} \sim a_0 M_{\mathrm{Pl}}^{1/2} H^{p + 1/2} \;,\;\;
\boxed{\text{term }04(2)} \;  \mathcal{R}_c^2 \partial^2 \mathcal{R}_c: \; E_{\text{str}} \sim a_0 M_{\mathrm{Pl}} H^{p}  \;,\\
\boxed{\text{term }05(1)} \;  &\mathcal{R}_c^{\prime}\left(\partial_i \mathcal{R}_c\right)^2: \; E_{\text{str}} \sim a_0 M_{\mathrm{Pl}}^{1/2} H^{p + 1/2}  \;,\;\; 
\boxed{\text{term }05(2)} \; \mathcal{R}_c\left(\partial_i \mathcal{R}_c\right)^2: \; E_{\text{str}} \sim a_0 M_{\mathrm{Pl}} H^{p}  \;,\\ 
\boxed{\text{term }06} \;  &\mathcal{R}_c\left(\partial_i \mathcal{R}_c\right)^2: \; E_{\text{str}} \sim a_0 M_{\mathrm{Pl}} H^{p}  \;,\;\; 
\boxed{\text{term }07(1)} \;  \mathcal{R}_c^{\prime}\left(\partial^2 \mathcal{R}_c\right)^2: \; E_{\text{str}} \sim a_0 M_{\mathrm{Pl}}^{1/4} H^{3/4+p} \;,
\end{aligned}
\end{equation*}

\begin{equation*}
\begin{aligned}
\boxed{\text{term }07(2)}\; &  \mathcal{R}_c\left(\partial^2 \mathcal{R}_c\right)^2: \; E_{\text{str}} \sim a_0 M_{\mathrm{Pl}}^{1/3} H^{2/3+p} \;,\;\;
\boxed{\text{term }08} \;  \mathcal{R}_c\left(\partial^2 \mathcal{R}_c\right)^2: \; E_{\text{str}} \sim a_0 M_{\mathrm{Pl}}^{1/3} H^{2/3+p} \;,\\
\boxed{\text{term }09} \; &  \partial^2 \mathcal{R}_c\left(\partial_i \mathcal{R}_c\right)^2: \; E_{\text{str}} \sim a_0 M_{\mathrm{Pl}}^{1/3} H^{2/3+p}  \;,\;\; \\
\boxed{\text{term }10(1)} \; &\mathcal{R}_c^{\prime}\left(\partial_i \partial_j \mathcal{R}_c\right)^2: \; E_{\text{str}} \sim a_0 M_{\mathrm{Pl}}^{1/4} H^{3/4+p}  \;,\\
\boxed{\text{term }10(2)} \; & \mathcal{R}_c\left(\partial_i \partial_j \mathcal{R}_c\right)^2: \; E_{\text{str}} \sim a_0 M_{\mathrm{Pl}}^{1/3} H^{2/3+p} \;,\;\;
\boxed{\text{term }11} \;  \mathcal{R}_c\left(\partial_i \partial_j \mathcal{R}_c\right)^2 \; E_{\text{str}} \sim a_0 M_{\mathrm{Pl}}^{1/3} H^{2/3+p}  \;,\\
\boxed{\text{term }12(1)} \; &  \mathcal{R}_c^{\prime} \partial_i \mathcal{R}_c \partial_i \psi_a: \; E_{\text{str}} \sim a_0 M_{\mathrm{Pl}} H^{p} \;,\;\; 
\boxed{\text{term }12(2)} \;   \mathcal{R}_c \partial_i \mathcal{R}_c \partial_i \psi_a: \; [ \mathcal{R}_c \partial_i \mathcal{R}_c \partial_i \psi_a] = 4\;,\\
\boxed{\text{term }12(3)} \; &  \mathcal{R}_c^{\prime} \partial_i \mathcal{R}_c \partial_i \partial^{-2} \mathcal{R}_c: \; [\mathcal{R}_c^{\prime} \partial_i \mathcal{R}_c \partial_i \partial^{-2} \mathcal{R}_c] = 4,\;
\\
\boxed{\text{term }12(4)} \;  &\mathcal{R}_c \partial_i \mathcal{R}_c \partial_i \partial^{-2} \mathcal{R}_c: \; [\mathcal{R}_c \partial_i \mathcal{R}_c \partial_i \partial^{-2} \mathcal{R}_c]=3 \;,\\ 
\boxed{\text{term }13(1)} \; & \partial^2 \mathcal{R}_c \partial_i \mathcal{R}_c \partial_i \psi_a: \; E_{\text{str}} \sim a_0 M_{\mathrm{Pl}}^{1/2} H^{p + 1/2}  \;,\;\; \\
\boxed{\text{term }13(2)} \;  &\partial^2 \mathcal{R}_c \partial_i \mathcal{R}_c \partial_i \partial^{-2} \mathcal{R}_c: \; E_{\text{str}} \sim a_0 M_{\mathrm{Pl}} H^{p}  \;,\\
\boxed{\text{term }14(1)} \; &  \mathcal{R}_c^{\prime}\left(\partial_i \partial_j \psi_a\right)^2: \; E_{\text{str}} \sim a_0 M_{\mathrm{Pl}}^{1/2} H^{p + 1/2} ,\,\\
\boxed{\text{term }14(2)} \,  &\mathcal{R}_c^{\prime}\left(\partial_i \partial_j \psi_a\right)\left(\partial_i \partial_j \partial^{-2} \mathcal{R}_c\right): \, E_{\text{str}} \sim a_0 M_{\mathrm{Pl}} H^{p}   ,\\
\boxed{\text{term }14(3)} \; & \mathcal{R}_c^{\prime}\left(\partial_i \partial_j \partial^{-2} \mathcal{R}_c\right)^2 \; [\mathcal{R}_c^{\prime}\left(\partial_i \partial_j \partial^{-2} \mathcal{R}_c\right)^2] = 4,\;
\boxed{\text{term }14(4)} \; \mathcal{R}_c\left(\partial_i \partial_j \psi_a\right)^2: \; E_{\text{str}} \sim a_0 M_{\mathrm{Pl}} H^{p}  \;,\\ 
\end{aligned}
\end{equation*}

\begin{equation*}
\begin{aligned}
\boxed{\text{term }14(5)} \; & \mathcal{R}_c\left(\partial_i \partial_j \psi_a\right)\left(\partial_i \partial_j \partial^{-2} \mathcal{R}_c\right) \; [\mathcal{R}_c\left(\partial_i \partial_j \psi_a\right)\left(\partial_i \partial_j \partial^{-2} \mathcal{R}_c\right) ] = 4,\;\\
\boxed{\text{term }14(6)} \; & \mathcal{R}_c\left(\partial_i \partial_j \partial^{-2} \mathcal{R}_c\right)^2 \; [\mathcal{R}_c\left(\partial_i \partial_j \partial^{-2} \mathcal{R}_c\right)^2]=3 ,\;  
\boxed{\text{term }15(1)} \;  \mathcal{R}_c\left(\partial_i \partial_j \psi_a\right)^2: \; E_{\text{str}} \sim a_0 M_{\mathrm{Pl}} H^{p} \;,\\
\boxed{\text{term }15(2)} \; &  \mathcal{R}_c\left(\partial_i \partial_j \psi_a\right)\left(\partial_i \partial_j \partial^{-2} \mathcal{R}_c\right) \; [\mathcal{R}_c\left(\partial_i \partial_j \psi_a\right)\left(\partial_i \partial_j \partial^{-2} \mathcal{R}_c\right)] = 4,\\
\boxed{\text{term }15(3)} \;  & \mathcal{R}_c\left(\partial_i \partial_j \partial^{-2} \mathcal{R}_c\right)^2: \; [\mathcal{R}_c\left(\partial_i \partial_j \partial^{-2} \mathcal{R}_c\right)^2]=3 \;,\\ 
\boxed{\text{term }16(1)} \;  &\mathcal{R}_c^{\prime}\left(\partial_i \partial_j \mathcal{R}_c\right) \partial_i \partial_j \psi_a: \;  E_{\text{str}} \sim a_0 M_{\mathrm{Pl}}^{1/3} H^{2/3+p}  \;,\\
\boxed{\text{term }16(2)}&\;  \mathcal{R}_c\left(\partial_i \partial_j \mathcal{R}_c\right) \partial_i \partial_j \psi_a: \; E_{\text{str}} \sim a_0 M_{\mathrm{Pl}}^{1/2} H^{p + 1/2}  \;,\\
\boxed{\text{term }16(3)} & \;  \mathcal{R}_c^{\prime}\left(\partial_i \partial_j \mathcal{R}_c\right) \partial_i \partial_j \partial^{-2} \mathcal{R}_c: \; E_{\text{str}} \sim a_0 M_{\mathrm{Pl}}^{1/2} H^{p + 1/2}  \;,\\ 
\boxed{\text{term }16(4)} \;&  \mathcal{R}_c\left(\partial_i \partial_j \mathcal{R}_c\right) \partial_i \partial_j \partial^{-2} \mathcal{R}_c: \; E_{\text{str}} \sim a_0 M_{\mathrm{Pl}} H^{p}  \;,\\
\boxed{\text{term }17(1)} \;& \mathcal{R}_c\left(\partial_i \partial_j \mathcal{R}_c\right)\left(\partial_i \partial_j \psi_a\right): \; E_{\text{str}} \sim a_0 M_{\mathrm{Pl}}^{1/2} H^{p + 1/2}  \;,\\
\boxed{\text{term }17(2)} \; &  \mathcal{R}_c\left(\partial_i \partial_j \mathcal{R}_c\right)\left(\partial_i \partial_j \partial^{-2} \mathcal{R}_c\right): \; E_{\text{str}} \sim a_0 M_{\mathrm{Pl}} H^{p}  \;. 
\end{aligned}
\end{equation*}
Terms with dimension $\leq 4$ are renormalizable and therefore do not lead to a strong-coupling scale. We conclude that the strongest unitarity bounds in the scalar sector arise from the interaction terms shown in Eq.\eqref{eq: L} and \eqref{eq: SL}.

\subsection{Mixed and Tensor Sectors}

In analogy with the scalar field case, we perform the canonical normalization for the tensor field:
\begin{equation}
h = a^{-1} \left(\sqrt{\mathcal{G}_T}\right)^{-1} h_c , \quad \dot{h} = a^{-1} \left(\sqrt{\mathcal{G}_T}\right)^{-1} \left( a^{-1} h_c^{\prime}- H h_c \right).
\end{equation}
Then the canonical quadratic action for tensor perturbations is given by:
\begin{equation}
\begin{aligned}
\mathcal{S}_{h h}^{(2)}
=&\frac{1}{8}\int \mathrm{~d} \eta \mathrm{~d}^3 x \left[ h^{\prime 2}_{c~ij} - c_T^2 h_{c~ij,k}^2 -m_{\mathrm{eff}} h_{c~ij}^2 \right], \quad m_{\mathrm{eff}}  \equiv-\frac{a''}{a} .
\end{aligned}
\end{equation}
While the canonically normalized cubic actions for the mixed sector are:
\begin{equation}
\begin{aligned}
\mathcal{L}^{(3)}_{\mathcal{R} h h} 
=& \left(\sqrt{2 \mathcal{G}_S}\right)^{-1} \left(\sqrt{ \mathcal{G}_T}\right)^{-2}  \\
\cdot& \left[d_1 \left(a^{-1} \mathcal{R}_c \left(h_{c~ij}^{\prime}\right)^2-2 H ~\mathcal{R}_c h_{c~ij}^{\prime} h_{c~ij}+ H^2 a~\mathcal{R}_c h_{c~ij}^2\right)\right.\\
& + d_2 a^{-1} \mathcal{R}_c h_{c~i j, k}^2\\
& \left.+d_3 \left(a^{-1} \psi_{c, k} h_{c~ i j}^{\prime} h_{c~i j, k} -H ~\psi_{c, k}h_{c~ i j} h_{c~i j, k} \right) \right] \\
& \left.+d_3 \left(-H ~\left(\partial _k\partial^{-2} \mathcal{R}_c\right) h_{c~ i j}^{\prime} h_{c~i j, k} +H^2 a~\left(\partial _k\partial^{-2} \mathcal{R}_c\right) h_{c~ i j} h_{c~i j, k} \right) \right] 
\end{aligned}
\end{equation}
and
\begin{equation}
\begin{aligned}
\mathcal{L}^{(3)}_{\mathcal{R} \mathcal{R} h}= & \left(\sqrt{2 \mathcal{G}_S}\right)^{-2} \left(\sqrt{ \mathcal{G}_T}\right)^{-1}  \\
\cdot& \left[ c_1 a^{-1} h_{c~i j} \mathcal{R}_{c, i} \mathcal{R}_{c, j}\right.\\
& + c_2 \left(a^{-2} h_{c~ij}^{\prime}\mathcal{R}_{c, i} \mathcal{R}_{c, j}-H a^{-1} h_{c~ij}\mathcal{R}_{c, i} \mathcal{R}_{c, j}\right) \\
& +c_3 \left(a^{-1} h_{c~ij}^{\prime}\mathcal{R}_{c, i}  \partial_j \psi_a-H ~h_{c~ij}^{\prime}\mathcal{R}_{c, i} \left(\partial_j \partial^{-2} \mathcal{R}_c\right) \right) \\
& +c_3 \left(-H ~h_{c~ij} \mathcal{R}_{c, i}  \partial_j \psi_a+H^2 a ~h_{c~ij} \mathcal{R}_{c, i} \left(\partial_j \partial^{-2} \mathcal{R}_c\right) \right) \\
& +c_6  \left(a^{-1} \left(\partial^2 h_{c~i j}\right)\partial_i \psi_a \partial_j \psi_a-H ~\left(\partial^2 h_{c~i j}\right)\partial_i \psi_a\left(\partial_j \partial^{-2} \mathcal{R}_c\right)\right) \\
& \left.+c_6 \left(-H~\left(\partial^2 h_{c~i j}\right)\left(\partial_i \partial^{-2} \mathcal{R}_c\right) \partial_j \psi_a+H^2 a~\left(\partial^2 h_{c~i j}\right)\left(\partial_i \partial^{-2} \mathcal{R}_c\right) \left(\partial_j \partial^{-2} \mathcal{R}_c\right)\right)\right] .
\end{aligned}
\end{equation}
The cubic action for the tensor sector is:
\begin{equation}
\begin{aligned}
\mathcal{L}^{(3)}_{h h h}
=&\frac{M_{\mathrm{Pl}}^{-1} }{4 a}\left(h_{c~i k} h_{c~j l}-\frac{1}{2} h_{c~i j} h_{c~k l}\right) h_{c~i j, k l} .
\end{aligned}
\end{equation}

Similarly, we estimate the strong-coupling energy scale for each term. \\
For $\mathcal{L}_{\mathcal{R} h h}^{(3)}$:
\begin{equation*}
\begin{aligned}
\boxed{\text{term }01(1)} \;  &\mathcal{R}_c\left(h_{c~ i j}^{\prime}\right)^2: E_{\text {str }} \sim a_0 M_{\mathrm{P l}} H^p\;,\;\;\boxed{\text{term }01(2)} \;  \mathcal{R}_c h_{c~ i j}^{\prime} h_{c~ i j}: [\mathcal{R}_c h_{c~ i j}^{\prime} h_{c~ i j}] = 4\;,\;\;\\
\boxed{\text{term }01(3)} \;  &\mathcal{R}_c h_{c~ i j}^2: [\mathcal{R}_c h_{c~ i j}^2]=3\;,\;\;\boxed{\text{term }02} \;  \mathcal{R}_c h_{c~ i j, k}^2: E_{\text {str }} \sim a_0 M_{\mathrm{P l}} H^p\;,\;\;\\
\boxed{\text{term }03(1)} \;  &\psi_{a, k} h_{c~ i j}^{\prime} h_{c~ i j, k}: E_{\text {str }} \sim a_0 M_{\mathrm{P l}} H^p\;,\;\;\boxed{\text{term }03(2)} \;  \psi_{a, k} h_{c~ i j} h_{c~ i j, k}: [\psi_{a, k} h_{c~ i j} h_{c~ i j, k}]=4\;,\;\;\\
\boxed{\text{term }03(3)} \;  &\left(\partial_k \partial^{-2} \mathcal{R}_c\right) h_{c~ i j}^{\prime} h_{c~ i j, k}: [\left(\partial_k \partial^{-2} \mathcal{R}_c\right) h_{c~ i j}^{\prime} h_{c~ i j, k}]=4\;,\;\;\\
\boxed{\text{term }03(4)} \;  &\left(\partial_k \partial^{-2} \mathcal{R}_c\right) h_{c~ i j} h_{c ~i j, k}: [\left(\partial_k \partial^{-2} \mathcal{R}_c\right) h_{c~ i j} h_{c ~i j, k}]=3\;.\;\;
\end{aligned}
\end{equation*}
For $\mathcal{L}_{\mathcal{R}\mathcal{R} h}^{(3)}$:
\begin{equation*}
\begin{aligned}
\boxed{\text{term }01} \;  &h_{c~ i j} \mathcal{R}_{c, i} \mathcal{R}_{c, j}: E_{\text {str }} \sim a_0 M_{\mathrm{P l}} H^p\;,\;\;\boxed{\text{term }02(1)} \;  h_{c~ i j}^{\prime} \mathcal{R}_{c, i} \mathcal{R}_{c, j}: E_{\text {str }} \sim a_0 M_{\mathrm{P l}}^{1 / 2} H^{p+1 / 2}\;,\;\;\\
\boxed{\text{term }02(2)} \;  &h_{c~ i j} \mathcal{R}_{c, i} \mathcal{R}_{c, j}: E_{\mathrm{str}} \sim a_0 M_{\mathrm{P l}} H^p\;,\;\;\boxed{\text{term }03(1)} \;  h'_{c~ i j}\mathcal{R}_{c, i} \partial_j \psi_a: E_{\text {str }} \sim a_0 M_{\mathrm{P l}} H^p\;,\;\;\\
\boxed{\text{term }03(2)} \;  &h'_{c~ i j}\mathcal{R}_{c, i}\left(\partial_j \partial^{-2} \mathcal{R}_c\right): [h'_{c~ i j}\mathcal{R}_{c, i}\left(\partial_j \partial^{-2} \mathcal{R}_c\right)]=4\;,\;\;\\
\boxed{\text{term }03(3)} \;  &h_{c~ i j} \mathcal{R}_{c, i} \partial_j \psi_a: [h_{c~ i j} \mathcal{R}_{c, i} \partial_j \psi_a]=4\;,\;\;\\
\boxed{\text{term }03(4)} \;  &h_{c~i j} \mathcal{R}_{c, i}\left(\partial_j \partial^{-2} \mathcal{R}_c\right): [h_{c~i j} \mathcal{R}_{c, i}\left(\partial_j \partial^{-2} \mathcal{R}_c\right)]=3\;,\;\;\\
\boxed{\text{term }06(1)} \;  &\left(\partial^2 h_{c~i j}\right) \partial_i \psi_a \partial_j \psi_a: E_{\text {str }} \sim a_0 M_{\mathrm{P l}} H^p\;,\;\;\\
\boxed{\text{term }06(2)} \;  &\left(\partial^2 h_{c~ i j}\right) \partial_i \psi_a\left(\partial_j \partial^{-2} \mathcal{R}_c\right): [\left(\partial^2 h_{c~ i j}\right) \partial_i \psi_a\left(\partial_j \partial^{-2} \mathcal{R}_c\right)]=4\;,\;\;\\
\boxed{\text{term }06(3)} \; 
&\left(\partial^2 h_{c~ i j}\right)\left(\partial_i \partial^{-2} \mathcal{R}_c\right) \partial_j \psi_a: [\left(\partial^2 h_{c~ i j}\right)\left(\partial_i \partial^{-2} \mathcal{R}_c\right) \partial_j \psi_a]=4\;,\;\;\\
\boxed{\text{term }06(4)} \; 
&\left(\partial^2 h_{c~ i j}\right)\left(\partial_i \partial^{-2} \mathcal{R}_c\right)\left(\partial_j \partial^{-2} \mathcal{R}_c\right): [\left(\partial^2 h_{c~ i j}\right)\left(\partial_i \partial^{-2} \mathcal{R}_c\right)\left(\partial_j \partial^{-2} \mathcal{R}_c\right)]=3\;.\;\;\\
\end{aligned}
\end{equation*}
For $\mathcal{L}_{hh h}^{(3)}$:
\begin{equation*}
\begin{aligned}
\boxed{\text{term }01} \;  &h_{c~ i k} h_{c~ j l} h_{c~ i j, k l}: E_{\text {str }} \sim a_0 M_{\mathrm{P l}} H^p\;,\;\;\boxed{\text{term }02} \;  h_{c~ i j} h_{c~ k l} h_{c~ i j, k l}: E_{\mathrm{str}} \sim a_0 M_{\mathrm{P l}} H^p\;.\;\;\\
\end{aligned}
\end{equation*}

All the cubic actions discussed above contribute at sub-leading order compared to the pure scalar sector.  Therefore, we conclude that the scalar cubic action provides the most significant contribution to the condition for the absence of strong coupling.

\section{Unitarity Bounds from Sub-Leading Terms}
\label{app: Calculating unitary bounds for sub-leading terms}
\numberwithin{equation}{section}

Here, we consider the sub-leading terms and compute the exact unitarity bounds.  
Namely, we examine the interaction action given in~Eq.\eqref{eq: SL}.
Let us begin with the $s$-channel.

\textbf{(1) $s$-channel:}  
We have
\[
p_1 + p_2 = (E_1, \vec{p}_1) + (E_2, \vec{p}_2) = (E, \vec{0})\, .
\]
A subtlety arises in the $s$-channel. For the term
\[
(2 \widetilde{\Lambda}_{16,(1)})
\bigl[
\mathcal{R}'_c(\partial_i \partial_j \mathcal{R}_c)\,
\partial_i \partial_j \partial^{-2}\mathcal{R}'_c
\bigr],
\]
one naively encounters a divergence of the form \(1/0^2\).  
To resolve this issue, we introduce a slight shift in the momentum values,
\begin{equation}
    \mathbf{p}_1 = \vec{p}_1, \qquad 
    \mathbf{p}_2 = \vec{p}_2 + \vec{\eta},
\end{equation}
and only afterward take the limit \( |\vec{\eta}| \to 0 \).  
Quantities in this shifted frame are denoted by boldface symbols.  
This procedure may be viewed as moving to a different frame together with a redefinition of momentum variables.

Next, we choose the direction of $\vec{p}_1$ along the $z$-axis,
\(
\mathbf{p}_1 = \vec{p}_1 = |\vec{p}_1|(0,0,1)
\).
The vector $\vec{\eta}$ is parameterized in spherical coordinates as
\begin{equation}
    \vec{\eta} = |\vec{\eta}| \bigl( \sin\theta,\ 0,\ \cos\theta \bigr).
\end{equation}
We choose $\vec{p}_3$ to form an angle $\vartheta$ with $\vec{p}_1$, with components
\begin{equation}
\vec{p}_3 = |\vec{p}_1|(\sin\vartheta,\ 0,\ \cos\vartheta)\, .
\end{equation}
Using energy and momentum conservation, boldface momenta give
\begin{align}
\mathbf{p}_3 &= \vec{p}_3 + \alpha\vec{\eta}, \quad 
\mathbf{p}_4 = \vec{p}_4 + (1-\alpha)\vec{\eta},\\[4pt]
|\mathbf{p}_2|^2 &= |\vec{p}_2+\vec{\eta}|^2 
= |\vec{p}_1|^2 - |\vec{p}_1|\,|\vec{\eta}|\,\cos\theta + |\vec{\eta}|^2,\\
|\mathbf{p}_3|^2 &= |\vec{p}_1|^2 + \alpha |\vec{p}_1|\,|\vec{\eta}|\, \cos (\theta-\vartheta)
+ \alpha^2 |\vec{\eta}|^2,\\
|\mathbf{p}_4|^2 &= |\vec{p}_1|^2 - (1-\alpha) |\vec{p}_1|\,|\vec{\eta}|\, \cos (\theta-\vartheta)
+ (1-\alpha)^2 |\vec{\eta}|^2,
\end{align}
where
\begin{equation}
\alpha = \frac{1}{2}
- \frac{\cos\theta}{2 \cos (\theta-\vartheta)}
+ O(|\vec{\eta}|)\;.
\end{equation}
The energies in new frame are
\begin{equation}
\begin{aligned}
\mathbf{E}_1 &= E_1,\\
\mathbf{E}_2 &= E_2 - c_S |\vec{\eta}| \cos\theta + o(|\vec{\eta}|),\\
\mathbf{E}_3 &= E_3 + \alpha c_S |\vec{\eta}| \cos (\theta-\vartheta) + o(|\vec{\eta}|),\\
\mathbf{E}_4 &= E_4 - (1-\alpha)c_S |\vec{\eta}| \cos (\theta-\vartheta) + o(|\vec{\eta}|).
\end{aligned}
\end{equation}
The naive singular contribution from the $\widetilde{\Lambda}_{16,(1)}$ term for the incoming legs would be
\begin{equation}
\begin{aligned}
&\bigl|
(2 \widetilde{\Lambda}_{16,(1)})
\bigl[
(-i\mathbf{E}_1)(i\mathbf{p}_{2,i})(i\mathbf{p}_{2,j})
(-i\eta_i)(-i\eta_j)(-i\vec{\eta})^{-2}(iE)
\\
&\qquad\quad
+ (-i\mathbf{E}_2)(i\mathbf{p}_{1,i})(i\mathbf{p}_{1,j})
(-i\eta_i)(-i\eta_j)(-i\vec{\eta})^{-2}(iE)
\bigr]
\bigr|
\\[4pt]
&= |-(2 \widetilde{\Lambda}_{16,(1)})| 
\frac{E^4}{4 c_S^2}\cos^2\theta  + O(|\vec{\eta}|)
\;\lesssim\;
|-(2 \widetilde{\Lambda}_{16,(1)})|
\frac{E^4}{4 c_S^2}\; .
\end{aligned}
\end{equation}
Its contribution is bounded between $-2 \widetilde{\Lambda}_{16,(1)} E^4 / (4 c_S^2)$ and zero.
For the outgoing legs, we have
\begin{equation}
\begin{aligned}
&\bigl|
(2 \widetilde{\Lambda}_{16,(1)})
\bigl[
(i\mathbf{E}_3)(-i\mathbf{p}_{4,i})(-i\mathbf{p}_{4,j})
(i\eta_i)(i\eta_j)(i\vec{\eta})^{-2}(-iE)
\\
&\qquad\quad
+ (i\mathbf{E}_4)(-i\mathbf{p}_{3,i})(-i\mathbf{p}_{3,j})
(i\eta_i)(i\eta_j)(i\vec{\eta})^{-2}(-iE)
\bigr]
\bigr|
\\[4pt]
&= |-(2 \widetilde{\Lambda}_{16,(1)})|
\frac{E^4}{4c_S^2}
\cos^2 (\theta-\vartheta)
+ O(|\vec{\eta}|)
\;\lesssim\;
|-(2 \widetilde{\Lambda}_{16,(1)})|
\frac{E^4}{4c_S^2}\; .
\end{aligned}
\end{equation}
Thus, the $s$-channel never truly diverges; however, the matrix element remains formally ambiguous. For the purposes of our analysis, these differences are negligible, being of order unity. 
The ambiguity in the matrix element arises from the presence of inverse spatial derivatives in the vertices, which appear after integrating out non-dynamical fields (i.e., constraints). One possible resolution is to reintroduce the Goldstone boson associated with the spontaneous breaking of time diffeomorphisms via a St\"uckelberg transformation (as done in Ref.~\cite{Cheung:2007st}), thereby restoring the broken symmetry and, in the decoupling limit, eliminating the spatially nonlocal terms in the action. 
The absolute magnitude of the contributions from these vertices is manifestly bounded by
\[
\left| -2 \widetilde{\Lambda}_{16,(1)} \right| \frac{E^4}{4 c_S^2}.
\]
Accordingly, we adopt the value that yields the strongest unitarity bounds. A complete treatment of this ambiguity, while potentially of academic interest, lies beyond the scope of the present work and is therefore left for future study.

Under these approximations, the $s$-channel matrix element becomes
\begin{equation}
\begin{aligned}
i\mathcal{M}_s
&= -i\,\frac{E^6}{4 c_S^4}
\left(2 \widetilde{\Lambda}_{16,(1)} + \widetilde{\Lambda}_{3,(1)}\right)^2 \,,
\end{aligned}
\end{equation}
when $\theta = 0$, while for $\theta = \pi/2$ we have
\begin{equation}
\begin{aligned}
i\mathcal{M}_s
&= -i\,\frac{E^6}{4 c_S^4}
\left(\widetilde{\Lambda}_{16,(1)} + \widetilde{\Lambda}_{3,(1)}\right)^2 \,.
\end{aligned}
\end{equation}
The $t$- and $u$-channels have no naive divergent terms, and therefore their calculations are straightforward.  


\textbf{(2) $t$-channel:}  
With $p_1-p_3=(E_1,\vec{p}_1)-(E_3,\vec{p}_3)=(0,\vec{p}_t)$,
\begin{equation}
\begin{aligned}
i\mathcal{M}_t
=& -i E^6 \frac{1-x}{32 c_S^6}
\left\{
2c_S^2\bigl((1-x)\widetilde{\Lambda}_{16,(1)}+\widetilde{\Lambda}_{3,(1)}\bigr)
+(1+x)\bigl(3(\widetilde{\Lambda}_{7,(2)}-\widetilde{\Lambda}_{8})+2\widetilde{\Lambda}_{9}\bigr)
\right\}^{\!2}.
\end{aligned}
\end{equation}

\textbf{(3) $u$-channel:}  
With $p_1-p_4=(E_1,\vec{p}_1)-(E_4,\vec{p}_4)=(0,\vec{p}_u)$,
\begin{equation}
\begin{aligned}
i\mathcal{M}_u
=& -i E^6 \frac{1+x}{32 c_S^6}
\left\{
2c_S^2\bigl((1+x)\widetilde{\Lambda}_{16,(1)}+\widetilde{\Lambda}_{3,(1)}\bigr)
+(1-x)\bigl(3(\widetilde{\Lambda}_{7,(2)}-\widetilde{\Lambda}_{8})+2\widetilde{\Lambda}_{9}\bigr)
\right\}^{\!2}.
\end{aligned}
\end{equation}
Therefore the total matrix element is 
\begin{equation}
\begin{aligned}
i\mathcal{M}_s + i\mathcal{M}_t + i\mathcal{M}_u
= -i\frac{E^6}{M_{\mathrm{Pl}}^2} H^{-(6 p+4)} \frac{a_0^{-6}}{2^{9} c_S^{6}} \cdot g(p,q)\,
\bigl[f_0(c_S,p,q) + x^2 f_2(c_S,p,q)\bigr]\;,
\end{aligned}
\end{equation}
where
\begin{equation}
\begin{aligned}
g \equiv &\, p^{6p+4}(q+3 p q)^4 \left[3 + (1+3 p)^2 q(-3 + 2 p(1+3 p) q) \right]^{-3},\\[2mm]
f_2 \equiv &
-(1-3 p(1+3 p) q)^2
\\
&+ c_S^2 \left[16(-1 + p(1+3 p) q)(-1 + 3 p(1+3 p) q)\right]
\\
&+ c_S^4 \Bigl[16 \left(3 + (1+3 p)^2 q(-3 + 2 p(1+3 p) q)\right) 
\left(-1 + q\left(-3 + p\left(-14 - 15 p + 2(1+3 p)^3 q\right)\right)\right)\Bigr].
\end{aligned}
\end{equation}
For $\theta = 0$,
\begin{align*}
f_0 \equiv &
(1-3 p(1+3 p) q)^2
\\
&+ c_S^2 \Bigl[ 16(1+3 p) q\bigl(-6 + 9 q + p\bigl(-10 + (73 + 2 p(98 + 87 p)) q 
\\
&- 4 (1+3 p)^3 (3 + 7 p) q^2 + 4 p(1+3 p)^5 q^3\bigr)\bigr)\Bigr]
\\
&+ c_S^4 \Bigl[ 64 (-1 + p(1+3 p) q)^2 \Bigr]\;,
\end{align*}
while for $\theta = \pi/2$,
\begin{equation}
\begin{aligned}
f_0 \equiv &
(1-3 p(1+3 p) q)^2
\\
&+ c_S^2 \left[16(-3 + p(1+3 p) q)(-1 + p(1+3 p) q)\right]
\\
&+ c_S^4 \left[64 (-1 + p(1+3 p) q)^2 \right]\;.
\end{aligned}
\end{equation}
In derivations of $g(p,q)$, $f_0(p,q)$ and $f_2(p,q)$, we have also used the expressions for $\widetilde{\Lambda}_i$, see Eq.\eqref{def:tildeLambda}.

We now calculate the partial-wave amplitudes (PWA).  
Since our model has a non-unity scalar sound speed, we use the general formulas of Ref.~\cite{Ageeva:2022byg}.  
The largest contribution comes from the $\ell=0$ mode. We therefore evaluate only $\widetilde{a}_0$ in  Eq.\eqref{def:PWA}:
\begin{equation}
\begin{aligned}
\widetilde{a}_0
&=\frac{1}{64 \pi c_S^3} \int_{-1}^{1} \mathrm{d}x\,\mathcal{M}
\\
&= - \frac{E^6}{M_{\mathrm{Pl}}^{2}}\,H^{-(6p+4)}\,\frac{a_0^{-6} }{2^{15}\pi c_S^{9}}\,g(p,q)\Bigl[2 f_0(c_S,p,q)+\frac{2}{3} f_2(c_S,p,q)\Bigr].
\end{aligned}
\end{equation}
Finally, requiring $|\mathrm{Re}[\widetilde{a}_0]| = 1/2$, one obtains the strong-coupling energy scale
\begin{equation}
\begin{aligned}
E_{\mathrm{strong}}
=
M_{\mathrm{Pl}}^{1/3} H^{\,p+2/3}\, 2^{7/3} \pi^{1/6} ~a_0 \left|
\, 
c_S^{-9}\,
g(p,q)\,
\left(2 f_0(c_S,p,q)+\frac{2}{3} f_2(c_S,p,q)\right)
\right|^{-1/6}.
\end{aligned}
\end{equation}
This is the result quoted in Section~\ref{sec: Exact Unitarity Bounds}.

\newpage

\bibliographystyle{JHEP}
\bibliography{ref}

@article{EPTA:2023fyk,
    author = "Antoniadis, J. and others",
    collaboration = "EPTA, InPTA:",
    title = "{The second data release from the European Pulsar Timing Array - III. Search for gravitational wave signals}",
    eprint = "2306.16214",
    archivePrefix = "arXiv",
    primaryClass = "astro-ph.HE",
    doi = "10.1051/0004-6361/202346844",
    journal = "Astron. Astrophys.",
    volume = "678",
    pages = "A50",
    year = "2023"
}

@misc{sasaki2025unveilingprimordialblackhole,
      title={Unveiling Primordial Black Hole Relics Through Induced Gravitational Waves}, 
      author={Misao Sasaki and Jianing Wang},
      year={2025},
      eprint={2512.22450},
      archivePrefix={arXiv},
      primaryClass={hep-ph},
      url={https://arxiv.org/abs/2512.22450}, 
}

@article{Pi:2024ert,
    author = "Pi, Shi and Sasaki, Misao and Takhistov, Volodymyr and Wang, Jianing",
    title = "{Primordial Black Hole formation from power spectrum with finite-width}",
    eprint = "2501.00295",
    archivePrefix = "arXiv",
    primaryClass = "astro-ph.CO",
    reportNumber = "YITP-24-184, KEK-QUP-2024-0028, KEK-TH-2676, KEK-Cosmo-0369",
    doi = "10.1088/1475-7516/2025/09/045",
    journal = "JCAP",
    volume = "09",
    pages = "045",
    year = "2025"
}

@article{Pi:2022zxs,
    author = "Pi, Shi and Wang, Jianing",
    title = "{Primordial black hole formation in Starobinsky's linear potential model}",
    eprint = "2209.14183",
    archivePrefix = "arXiv",
    primaryClass = "astro-ph.CO",
    reportNumber = "IPMU22-0047",
    doi = "10.1088/1475-7516/2023/06/018",
    journal = "JCAP",
    volume = "06",
    pages = "018",
    year = "2023"
}

@article{Zhu:2017jew,
    author = "Zhu, Tao and Wang, Anzhong and Cleaver, Gerald and Kirsten, Klaus and Sheng, Qin",
    title = "{Pre-inflationary universe in loop quantum cosmology}",
    eprint = "1705.07544",
    archivePrefix = "arXiv",
    primaryClass = "gr-qc",
    doi = "10.1103/PhysRevD.96.083520",
    journal = "Phys. Rev. D",
    volume = "96",
    number = "8",
    pages = "083520",
    year = "2017"
}

@article{Powell:2006yg,
    author = "Powell, Brian A. and Kinney, William H.",
    title = "{The pre-inflationary vacuum in the cosmic microwave background}",
    eprint = "astro-ph/0612006",
    archivePrefix = "arXiv",
    doi = "10.1103/PhysRevD.76.063512",
    journal = "Phys. Rev. D",
    volume = "76",
    pages = "063512",
    year = "2007"
}

@article{Bahrami:2015bva,
    author = "Bahrami, Sina and Flanagan, Eanna E.",
    title = "{Sensitivity of inflationary predictions to pre-inflationary phases}",
    eprint = "1505.00745",
    archivePrefix = "arXiv",
    primaryClass = "hep-th",
    doi = "10.1088/1475-7516/2016/01/027",
    journal = "JCAP",
    volume = "01",
    pages = "027",
    year = "2016"
}

@article{NANOGrav:2023gor,
    author = "Agazie, Gabriella and others",
    collaboration = "NANOGrav",
    title = "{The NANOGrav 15 yr Data Set: Evidence for a Gravitational-wave Background}",
    eprint = "2306.16213",
    archivePrefix = "arXiv",
    primaryClass = "astro-ph.HE",
    doi = "10.3847/2041-8213/acdac6",
    journal = "Astrophys. J. Lett.",
    volume = "951",
    number = "1",
    pages = "L8",
    year = "2023"
}

@article{Franciolini:2023pbf,
    author = "Franciolini, Gabriele and Iovino, Junior., Antonio and Vaskonen, Ville and Veermae, Hardi",
    title = "{Recent Gravitational Wave Observation by Pulsar Timing Arrays and Primordial Black Holes: The Importance of Non-Gaussianities}",
    eprint = "2306.17149",
    archivePrefix = "arXiv",
    primaryClass = "astro-ph.CO",
    doi = "10.1103/PhysRevLett.131.201401",
    journal = "Phys. Rev. Lett.",
    volume = "131",
    number = "20",
    pages = "201401",
    year = "2023"
}

@article{Jiang:2024dxj,
    author = "Jiang, Jun-Qian and Piao, Yun-Song",
    title = "{Search for the non-linearities of gravitational wave background in NANOGrav 15-year data set}",
    eprint = "2401.16950",
    archivePrefix = "arXiv",
    primaryClass = "gr-qc",
    doi = "10.1016/j.physletb.2025.139284",
    journal = "Phys. Lett. B",
    volume = "862",
    pages = "139284",
    year = "2025"
}

@article{Domenech:2024rks,
    author = "Dom{\`e}nech, Guillem and Pi, Shi and Wang, Ao and Wang, Jianing",
    title = "{Induced gravitational wave interpretation of PTA data: a complete study for general equation of state}",
    eprint = "2402.18965",
    archivePrefix = "arXiv",
    primaryClass = "astro-ph.CO",
    doi = "10.1088/1475-7516/2024/08/054",
    journal = "JCAP",
    volume = "08",
    pages = "054",
    year = "2024"
}

@article{Tan:2024kuk,
    author = "Tan, Qin and Wu, You and Liu, Lang",
    title = "{Constraining string cosmology with the gravitational-wave background using the NANOGrav 15-year data set}",
    eprint = "2409.17846",
    archivePrefix = "arXiv",
    primaryClass = "gr-qc",
    doi = "10.1140/epjc/s10052-025-13998-1",
    journal = "Eur. Phys. J. C",
    volume = "85",
    number = "3",
    pages = "327",
    year = "2025"
}

@article{Kenjale:2024rsc,
    author = "Kenjale, Ved and Kahniashvili, Tina",
    title = "{Connecting inflation to the NANOGrav 15-year dataset via massive gravity}",
    eprint = "2410.09658",
    archivePrefix = "arXiv",
    primaryClass = "astro-ph.CO",
    doi = "10.1103/PhysRevD.111.103515",
    journal = "Phys. Rev. D",
    volume = "111",
    number = "10",
    pages = "103515",
    year = "2025"
}

@article{Goodhew:2020hob,
    author = "Goodhew, Harry and Jazayeri, Sadra and Pajer, Enrico",
    title = "{The Cosmological Optical Theorem}",
    eprint = "2009.02898",
    archivePrefix = "arXiv",
    primaryClass = "hep-th",
    doi = "10.1088/1475-7516/2021/04/021",
    journal = "JCAP",
    volume = "04",
    pages = "021",
    year = "2021"
}

@article{Kobayashi:2025jkg,
    author = "Kobayashi, Mikage U. and Kohri, Kazunori",
    title = "{NANOGrav 15-year gravitational-wave signals from binary supermassive black-holes seeded by primordial black holes}",
    eprint = "2511.04210",
    archivePrefix = "arXiv",
    primaryClass = "astro-ph.CO",
    reportNumber = "KEK-TH-2776, KEK-Cosmo-0397",
    month = "11",
    year = "2025"
}

@article{Tahara:2020fmn,
    author = "Tahara, Hiroaki W. H. and Kobayashi, Tsutomu",
    title = "{Nanohertz gravitational waves from a null-energy-condition violation in the early universe}",
    eprint = "2011.01605",
    archivePrefix = "arXiv",
    primaryClass = "gr-qc",
    reportNumber = "RUP-20-32",
    doi = "10.1103/PhysRevD.102.123533",
    journal = "Phys. Rev. D",
    volume = "102",
    number = "12",
    pages = "123533",
    year = "2020"
}

@article{Borah:2023sbc,
    author = "Borah, Debasish and Jyoti Das, Suruj and Samanta, Rome",
    title = "{Imprint of inflationary gravitational waves and WIMP dark matter in pulsar timing array data}",
    eprint = "2307.00537",
    archivePrefix = "arXiv",
    primaryClass = "hep-ph",
    doi = "10.1088/1475-7516/2024/03/031",
    journal = "JCAP",
    volume = "03",
    pages = "031",
    year = "2024"
}

@article{Zhu:2023lbf,
    author = "Zhu, Mian and Ye, Gen and Cai, Yong",
    title = "{Pulsar timing array observations as possible hints for nonsingular cosmology}",
    eprint = "2307.16211",
    archivePrefix = "arXiv",
    primaryClass = "astro-ph.CO",
    doi = "10.1140/epjc/s10052-023-11963-4",
    journal = "Eur. Phys. J. C",
    volume = "83",
    number = "9",
    pages = "816",
    year = "2023"
}

@article{Vagnozzi:2023lwo,
    author = "Vagnozzi, Sunny",
    title = "{Inflationary interpretation of the stochastic gravitational wave background signal detected by pulsar timing array experiments}",
    eprint = "2306.16912",
    archivePrefix = "arXiv",
    primaryClass = "astro-ph.CO",
    doi = "10.1016/j.jheap.2023.07.001",
    journal = "JHEAp",
    volume = "39",
    pages = "81--98",
    year = "2023"
}

@article{Ye:2023tpz,
    author = "Ye, Gen and Zhu, Mian and Cai, Yong",
    title = "{Null energy condition violation during inflation and pulsar timing array observations}",
    eprint = "2312.10685",
    archivePrefix = "arXiv",
    primaryClass = "gr-qc",
    doi = "10.1007/JHEP02(2024)008",
    journal = "JHEP",
    volume = "02",
    pages = "008",
    year = "2024"
}

@article{Chen:2024jca,
    author = "Chen, Zu-Cheng and Liu, Lang",
    title = "{Detecting a gravitational wave background from inflation with null energy condition violation: prospects for Taiji}",
    eprint = "2404.08375",
    archivePrefix = "arXiv",
    primaryClass = "gr-qc",
    doi = "10.1140/epjc/s10052-024-13529-4",
    journal = "Eur. Phys. J. C",
    volume = "84",
    number = "11",
    pages = "1176",
    year = "2024"
}

@article{Akama:2024vgu,
    author = "Akama, Shingo and Hirano, Shin'ichi and Yokoyama, Shuichiro",
    title = "{Stochastic gravitational wave background anisotropies from inflation with non-Bunch-Davies states}",
    eprint = "2410.14664",
    archivePrefix = "arXiv",
    primaryClass = "astro-ph.CO",
    reportNumber = "RUP-24-20",
    doi = "10.1093/ptep/ptaf117",
    month = "10",
    year = "2024"
}

@article{Rubakov:2014jja,
    author = "Rubakov, V. A.",
    title = "{The Null Energy Condition and its violation}",
    eprint = "1401.4024",
    archivePrefix = "arXiv",
    primaryClass = "hep-th",
    reportNumber = "INR-TH-2014-1",
    doi = "10.3367/UFNe.0184.201402b.0137",
    journal = "Phys. Usp.",
    volume = "57",
    pages = "128--142",
    year = "2014"
}

@article{Horndeski:1974wa,
    author = "Horndeski, Gregory Walter",
    title = "{Second-order scalar-tensor field equations in a four-dimensional space}",
    doi = "10.1007/BF01807638",
    journal = "Int. J. Theor. Phys.",
    volume = "10",
    pages = "363--384",
    year = "1974"
}

@article{Zumalacarregui:2013pma,
    author = "Zumalac{\'a}rregui, Miguel and Garc{\'\i}a-Bellido, Juan",
    title = "{Transforming gravity: from derivative couplings to matter to second-order scalar-tensor theories beyond the Horndeski Lagrangian}",
    eprint = "1308.4685",
    archivePrefix = "arXiv",
    primaryClass = "gr-qc",
    reportNumber = "IFT-UAM-CSIC-13-090",
    doi = "10.1103/PhysRevD.89.064046",
    journal = "Phys. Rev. D",
    volume = "89",
    pages = "064046",
    year = "2014"
}

@article{Gleyzes:2014dya,
    author = "Gleyzes, J{\'e}r{\^o}me and Langlois, David and Piazza, Federico and Vernizzi, Filippo",
    title = "{Healthy theories beyond Horndeski}",
    eprint = "1404.6495",
    archivePrefix = "arXiv",
    primaryClass = "hep-th",
    doi = "10.1103/PhysRevLett.114.211101",
    journal = "Phys. Rev. Lett.",
    volume = "114",
    number = "21",
    pages = "211101",
    year = "2015"
}

@article{Langlois:2015cwa,
    author = "Langlois, David and Noui, Karim",
    title = "{Degenerate higher derivative theories beyond Horndeski: evading the Ostrogradski instability}",
    eprint = "1510.06930",
    archivePrefix = "arXiv",
    primaryClass = "gr-qc",
    doi = "10.1088/1475-7516/2016/02/034",
    journal = "JCAP",
    volume = "02",
    pages = "034",
    year = "2016"
}

@article{Kobayashi:2019hrl,
    author = "Kobayashi, Tsutomu",
    title = "{Horndeski theory and beyond: a review}",
    eprint = "1901.07183",
    archivePrefix = "arXiv",
    primaryClass = "gr-qc",
    reportNumber = "RUP-19-3",
    doi = "10.1088/1361-6633/ab2429",
    journal = "Rept. Prog. Phys.",
    volume = "82",
    number = "8",
    pages = "086901",
    year = "2019"
}

@article{Creminelli:2010ba,
    author = "Creminelli, Paolo and Nicolis, Alberto and Trincherini, Enrico",
    title = "{Galilean Genesis: An Alternative to inflation}",
    eprint = "1007.0027",
    archivePrefix = "arXiv",
    primaryClass = "hep-th",
    doi = "10.1088/1475-7516/2010/11/021",
    journal = "JCAP",
    volume = "11",
    pages = "021",
    year = "2010"
}

@article{Creminelli:2012my,
    author = "Creminelli, Paolo and Hinterbichler, Kurt and Khoury, Justin and Nicolis, Alberto and Trincherini, Enrico",
    title = "{Subluminal Galilean Genesis}",
    eprint = "1209.3768",
    archivePrefix = "arXiv",
    primaryClass = "hep-th",
    doi = "10.1007/JHEP02(2013)006",
    journal = "JHEP",
    volume = "02",
    pages = "006",
    year = "2013"
}

@article{Hinterbichler:2012fr,
    author = "Hinterbichler, Kurt and Joyce, Austin and Khoury, Justin and Miller, Godfrey E. J.",
    title = "{DBI Realizations of the Pseudo-Conformal Universe and Galilean Genesis Scenarios}",
    eprint = "1209.5742",
    archivePrefix = "arXiv",
    primaryClass = "hep-th",
    doi = "10.1088/1475-7516/2012/12/030",
    journal = "JCAP",
    volume = "12",
    pages = "030",
    year = "2012"
}

@article{Elder:2013gya,
    author = "Elder, Benjamin and Joyce, Austin and Khoury, Justin",
    title = "{From Satisfying to Violating the Null Energy Condition}",
    eprint = "1311.5889",
    archivePrefix = "arXiv",
    primaryClass = "hep-th",
    doi = "10.1103/PhysRevD.89.044027",
    journal = "Phys. Rev. D",
    volume = "89",
    number = "4",
    pages = "044027",
    year = "2014"
}

@article{Pirtskhalava:2014esa,
    author = "Pirtskhalava, David and Santoni, Luca and Trincherini, Enrico and Uttayarat, Patipan",
    title = "{Inflation from Minkowski Space}",
    eprint = "1410.0882",
    archivePrefix = "arXiv",
    primaryClass = "hep-th",
    doi = "10.1007/JHEP12(2014)151",
    journal = "JHEP",
    volume = "12",
    pages = "151",
    year = "2014"
}

@article{Nishi:2015pta,
    author = "Nishi, Sakine and Kobayashi, Tsutomu",
    title = "{Generalized Galilean Genesis}",
    eprint = "1501.02553",
    archivePrefix = "arXiv",
    primaryClass = "hep-th",
    reportNumber = "RUP-15-1",
    doi = "10.1088/1475-7516/2015/03/057",
    journal = "JCAP",
    volume = "03",
    pages = "057",
    year = "2015"
}

@article{Kobayashi:2015gga,
    author = "Kobayashi, Tsutomu and Yamaguchi, Masahide and Yokoyama, Jun'ichi",
    title = "{Galilean Creation of the Inflationary Universe}",
    eprint = "1504.05710",
    archivePrefix = "arXiv",
    primaryClass = "hep-th",
    reportNumber = "RUP-15-6, RESCEU-10-15",
    doi = "10.1088/1475-7516/2015/07/017",
    journal = "JCAP",
    volume = "07",
    pages = "017",
    year = "2015"
}

@article{Ilyas:2020zcb,
    author = "Ilyas, Amara and Zhu, Mian and Zheng, Yunlong and Cai, Yi-Fu",
    title = "{Emergent Universe and Genesis from the DHOST Cosmology}",
    eprint = "2009.10351",
    archivePrefix = "arXiv",
    primaryClass = "gr-qc",
    doi = "10.1007/JHEP01(2021)141",
    journal = "JHEP",
    volume = "01",
    pages = "141",
    year = "2021"
}

@article{Zhu:2021ggm,
    author = "Zhu, Mian and Zheng, Yunlong",
    title = "{Improved DHOST Genesis}",
    eprint = "2109.05277",
    archivePrefix = "arXiv",
    primaryClass = "gr-qc",
    doi = "10.1007/JHEP11(2021)163",
    journal = "JHEP",
    volume = "11",
    pages = "163",
    year = "2021"
}

@article{Qiu:2011cy,
    author = "Qiu, Taotao and Evslin, Jarah and Cai, Yi-Fu and Li, Mingzhe and Zhang, Xinmin",
    title = "{Bouncing Galileon Cosmologies}",
    eprint = "1108.0593",
    archivePrefix = "arXiv",
    primaryClass = "hep-th",
    doi = "10.1088/1475-7516/2011/10/036",
    journal = "JCAP",
    volume = "10",
    pages = "036",
    year = "2011"
}

@article{Easson:2011zy,
    author = "Easson, Damien A. and Sawicki, Ignacy and Vikman, Alexander",
    title = "{G-Bounce}",
    eprint = "1109.1047",
    archivePrefix = "arXiv",
    primaryClass = "hep-th",
    reportNumber = "CERN-PH-TH-2011-203",
    doi = "10.1088/1475-7516/2011/11/021",
    journal = "JCAP",
    volume = "11",
    pages = "021",
    year = "2011"
}

@article{Battarra:2014tga,
    author = "Battarra, Lorenzo and Koehn, Michael and Lehners, Jean-Luc and Ovrut, Burt A.",
    title = "{Cosmological Perturbations Through a Non-Singular Ghost-Condensate/Galileon Bounce}",
    eprint = "1404.5067",
    archivePrefix = "arXiv",
    primaryClass = "hep-th",
    doi = "10.1088/1475-7516/2014/07/007",
    journal = "JCAP",
    volume = "07",
    pages = "007",
    year = "2014"
}

@article{Nicolis:2009qm,
    author = "Nicolis, Alberto and Rattazzi, Riccardo and Trincherini, Enrico",
    title = "{Energy's and amplitudes' positivity}",
    eprint = "0912.4258",
    archivePrefix = "arXiv",
    primaryClass = "hep-th",
    doi = "10.1007/JHEP05(2010)095",
    journal = "JHEP",
    volume = "05",
    pages = "095",
    year = "2010",
    note = "[Erratum: JHEP 11, 128 (2011)]"
}

@article{Ageeva:2018lko,
    author = "Ageeva, Y. A. and Evseev, O. A. and Melichev, O. I. and Rubakov, V. A.",
    editor = "Volkova, V. E. and Zhezher, Y. V. and Levkov, D. G. and Rubakov, V. A. and Matveev, V. A.",
    title = "{Horndeski Genesis: strong coupling and absence thereof}",
    eprint = "1810.00465",
    archivePrefix = "arXiv",
    primaryClass = "hep-th",
    reportNumber = "INR-TH-2018-025",
    doi = "10.1051/epjconf/201819107010",
    journal = "EPJ Web Conf.",
    volume = "191",
    pages = "07010",
    year = "2018"
}

@article{Ageeva:2020gti,
    author = "Ageeva, Y. and Evseev, O. and Melichev, O. and Rubakov, V.",
    title = "{Toward evading the strong coupling problem in Horndeski genesis}",
    eprint = "2003.01202",
    archivePrefix = "arXiv",
    primaryClass = "hep-th",
    reportNumber = "INR-TH-2020-006",
    doi = "10.1103/PhysRevD.102.023519",
    journal = "Phys. Rev. D",
    volume = "102",
    number = "2",
    pages = "023519",
    year = "2020"
}

@article{Ageeva:2020buc,
    author = "Ageeva, Yulia and Petrov, Pavel and Rubakov, Valery",
    title = "{Horndeski genesis: consistency of classical theory}",
    eprint = "2009.05071",
    archivePrefix = "arXiv",
    primaryClass = "hep-th",
    reportNumber = "INR-TH-2020-038",
    doi = "10.1007/JHEP12(2020)107",
    journal = "JHEP",
    volume = "12",
    pages = "107",
    year = "2020"
}

@article{Ageeva:2022fyq,
    author = "Ageeva, Yulia and Petrov, Pavel",
    title = "{On the strong coupling problem in cosmologies with {\textquotedblleft}strong gravity in the past{\textquotedblright}}",
    eprint = "2206.10646",
    archivePrefix = "arXiv",
    primaryClass = "gr-qc",
    reportNumber = "INR-TH-2022-013",
    doi = "10.1142/S0217732322501711",
    journal = "Mod. Phys. Lett. A",
    volume = "37",
    number = "26",
    pages = "2250171",
    year = "2022"
}

@article{Cai:2022ori,
    author = "Cai, Yong and Xu, Ji and Zhao, Shuai and Zhou, Siyi",
    title = "{Perturbative unitarity and NEC violation in genesis cosmology}",
    eprint = "2207.11772",
    archivePrefix = "arXiv",
    primaryClass = "gr-qc",
    reportNumber = "JLAB-THY-22-3668",
    doi = "10.1007/JHEP10(2022)140",
    journal = "JHEP",
    volume = "10",
    pages = "140",
    year = "2022",
    note = "[Erratum: JHEP 11, 063 (2022)]"
}

@article{Ageeva:2022asq,
    author = "Ageeva, Yulia and Petrov, Pavel and Rubakov, Valery",
    title = "{Generating cosmological perturbations in non-singular Horndeski cosmologies}",
    eprint = "2207.04071",
    archivePrefix = "arXiv",
    primaryClass = "hep-th",
    reportNumber = "INR-TH-2022-014",
    doi = "10.1007/JHEP01(2023)026",
    journal = "JHEP",
    volume = "01",
    pages = "026",
    year = "2023"
}

@article{Ageeva:2023nwf,
    author = "Ageeva, Y. and Petrov, P.",
    title = "{K-inflation: The legitimacy of the classical treatment}",
    eprint = "2310.18402",
    archivePrefix = "arXiv",
    primaryClass = "hep-th",
    reportNumber = "INR-TH-2023-019",
    doi = "10.1103/PhysRevD.110.043527",
    journal = "Phys. Rev. D",
    volume = "110",
    number = "4",
    pages = "043527",
    year = "2024"
}

@article{Baumann:2011dt,
    author = "Baumann, Daniel and Senatore, Leonardo and Zaldarriaga, Matias",
    title = "{Scale-Invariance and the Strong Coupling Problem}",
    eprint = "1101.3320",
    archivePrefix = "arXiv",
    primaryClass = "hep-th",
    doi = "10.1088/1475-7516/2011/05/004",
    journal = "JCAP",
    volume = "05",
    pages = "004",
    year = "2011"
}

@article{Bueno:2016xff,
    author = "Bueno, Pablo and Cano, Pablo A.",
    title = "{Einsteinian cubic gravity}",
    eprint = "1607.06463",
    archivePrefix = "arXiv",
    primaryClass = "hep-th",
    doi = "10.1103/PhysRevD.94.104005",
    journal = "Phys. Rev. D",
    volume = "94",
    number = "10",
    pages = "104005",
    year = "2016"
}

@article{Hu:2023juh,
    author = "Hu, Yu-Min and Zhao, Yaqi and Ren, Xin and Wang, Bo and Saridakis, Emmanuel N. and Cai, Yi-Fu",
    title = "{The effective field theory approach to the strong coupling issue in f(T) gravity}",
    eprint = "2302.03545",
    archivePrefix = "arXiv",
    primaryClass = "gr-qc",
    doi = "10.1088/1475-7516/2023/07/060",
    journal = "JCAP",
    volume = "07",
    pages = "060",
    year = "2023"
}

@article{Cannone:2014qna,
    author = "Cannone, Dario and Bartolo, Nicola and Matarrese, Sabino",
    title = "{Perturbative Unitarity of Inflationary Models with Features}",
    eprint = "1402.2258",
    archivePrefix = "arXiv",
    primaryClass = "astro-ph.CO",
    doi = "10.1103/PhysRevD.89.127301",
    journal = "Phys. Rev. D",
    volume = "89",
    number = "12",
    pages = "127301",
    year = "2014"
}

@article{Escriva:2016cwl,
    author = "Escriv{\`a}, Albert and Germani, Cristiano",
    title = "{Beyond dimensional analysis: Higgs and new Higgs inflations do not violate unitarity}",
    eprint = "1612.06253",
    archivePrefix = "arXiv",
    primaryClass = "hep-ph",
    reportNumber = "ICCUB-16-043",
    doi = "10.1103/PhysRevD.95.123526",
    journal = "Phys. Rev. D",
    volume = "95",
    number = "12",
    pages = "123526",
    year = "2017"
}

@article{deRham:2017avq,
    author = "de Rham, Claudia and Melville, Scott and Tolley, Andrew J. and Zhou, Shuang-Yong",
    title = "{Positivity bounds for scalar field theories}",
    eprint = "1702.06134",
    archivePrefix = "arXiv",
    primaryClass = "hep-th",
    doi = "10.1103/PhysRevD.96.081702",
    journal = "Phys. Rev. D",
    volume = "96",
    number = "8",
    pages = "081702",
    year = "2017"
}

@article{Kim:2021pbr,
    author = "Kim, Suro and Noumi, Toshifumi and Takeuchi, Keito and Zhou, Siyi",
    title = "{Perturbative unitarity in quasi-single field inflation}",
    eprint = "2102.04101",
    archivePrefix = "arXiv",
    primaryClass = "hep-th",
    reportNumber = "KOBE-COSMO-21-01",
    doi = "10.1007/JHEP07(2021)018",
    journal = "JHEP",
    volume = "07",
    pages = "018",
    year = "2021"
}

@article{GilChoi:2025hbs,
    author = "Gil Choi, Han and Petrov, Pavel and Yamaguchi, Masahide",
    title = "{Can Horndeski Genesis be nonpathological?}",
    eprint = "2503.02626",
    archivePrefix = "arXiv",
    primaryClass = "hep-th",
    doi = "10.1007/JHEP08(2025)044",
    journal = "JHEP",
    volume = "08",
    pages = "044",
    year = "2025"
}

@article{deRham:2017aoj,
    author = "de Rham, Claudia and Melville, Scott",
    title = "{Unitary null energy condition violation in P(X) cosmologies}",
    eprint = "1703.00025",
    archivePrefix = "arXiv",
    primaryClass = "hep-th",
    doi = "10.1103/PhysRevD.95.123523",
    journal = "Phys. Rev. D",
    volume = "95",
    number = "12",
    pages = "123523",
    year = "2017"
}

@article{Ageeva:2022byg,
    author = "Ageeva, Yuliya Aleksandrovna and Petrov, Pavel Konstantinovich",
    title = "{Unitarity relation and unitarity bounds for scalars with different sound speeds}",
    eprint = "2206.03516",
    archivePrefix = "arXiv",
    primaryClass = "hep-th",
    reportNumber = "INR-TH-2022-011",
    doi = "10.3367/UFNe.2022.11.039259",
    journal = "Phys. Usp.",
    volume = "66",
    number = "11",
    pages = "1134--1141",
    year = "2023"
}

@article{Libanov:2016kfc,
    author = "Libanov, M. and Mironov, S. and Rubakov, V.",
    title = "{Generalized Galileons: instabilities of bouncing and Genesis cosmologies and modified Genesis}",
    eprint = "1605.05992",
    archivePrefix = "arXiv",
    primaryClass = "hep-th",
    reportNumber = "INR-TH-2016-014",
    doi = "10.1088/1475-7516/2016/08/037",
    journal = "JCAP",
    volume = "08",
    pages = "037",
    year = "2016"
}

@article{Creminelli:2016zwa,
    author = "Creminelli, Paolo and Pirtskhalava, David and Santoni, Luca and Trincherini, Enrico",
    title = "{Stability of Geodesically Complete Cosmologies}",
    eprint = "1610.04207",
    archivePrefix = "arXiv",
    primaryClass = "hep-th",
    doi = "10.1088/1475-7516/2016/11/047",
    journal = "JCAP",
    volume = "11",
    pages = "047",
    year = "2016"
}

@article{Kobayashi:2016xpl,
    author = "Kobayashi, Tsutomu",
    title = "{Generic instabilities of nonsingular cosmologies in Horndeski theory: A no-go theorem}",
    eprint = "1606.05831",
    archivePrefix = "arXiv",
    primaryClass = "hep-th",
    reportNumber = "RUP-16-19",
    doi = "10.1103/PhysRevD.94.043511",
    journal = "Phys. Rev. D",
    volume = "94",
    number = "4",
    pages = "043511",
    year = "2016"
}

@article{Kobayashi:2011nu,
    author = "Kobayashi, Tsutomu and Yamaguchi, Masahide and Yokoyama, Jun'ichi",
    title = "{Generalized G-inflation: Inflation with the most general second-order field equations}",
    eprint = "1105.5723",
    archivePrefix = "arXiv",
    primaryClass = "hep-th",
    reportNumber = "KUNS-2339, RESCEU-9-11",
    doi = "10.1143/PTP.126.511",
    journal = "Prog. Theor. Phys.",
    volume = "126",
    pages = "511--529",
    year = "2011"
}

@article{Gao:2011qe,
    author = "Gao, Xian and Steer, Daniele A.",
    title = "{Inflation and primordial non-Gaussianities of 'generalized Galileons'}",
    eprint = "1107.2642",
    archivePrefix = "arXiv",
    primaryClass = "astro-ph.CO",
    doi = "10.1088/1475-7516/2011/12/019",
    journal = "JCAP",
    volume = "12",
    pages = "019",
    year = "2011"
}

@article{Gao:2011vs,
    author = "Gao, Xian and Kobayashi, Tsutomu and Yamaguchi, Masahide and Yokoyama, Jun'ichi",
    title = "{Primordial non-Gaussianities of gravitational waves in the most general single-field inflation model}",
    eprint = "1108.3513",
    archivePrefix = "arXiv",
    primaryClass = "astro-ph.CO",
    doi = "10.1103/PhysRevLett.107.211301",
    journal = "Phys. Rev. Lett.",
    volume = "107",
    pages = "211301",
    year = "2011"
}

@article{DeFelice:2010nf,
    author = "De Felice, Antonio and Tsujikawa, Shinji",
    title = "{Generalized Galileon cosmology}",
    eprint = "1008.4236",
    archivePrefix = "arXiv",
    primaryClass = "hep-th",
    doi = "10.1103/PhysRevD.84.124029",
    journal = "Phys. Rev. D",
    volume = "84",
    pages = "124029",
    year = "2011"
}

@article{DeFelice:2011uc,
    author = "De Felice, Antonio and Tsujikawa, Shinji",
    title = "{Inflationary non-Gaussianities in the most general second-order scalar-tensor theories}",
    eprint = "1107.3917",
    archivePrefix = "arXiv",
    primaryClass = "gr-qc",
    doi = "10.1103/PhysRevD.84.083504",
    journal = "Phys. Rev. D",
    volume = "84",
    pages = "083504",
    year = "2011"
}

@article{Renaux-Petel:2011zgy,
    author = "Renaux-Petel, Sebastien",
    title = "{On the redundancy of operators and the bispectrum in the most general second-order scalar-tensor theory}",
    eprint = "1107.5020",
    archivePrefix = "arXiv",
    primaryClass = "astro-ph.CO",
    doi = "10.1088/1475-7516/2012/02/020",
    journal = "JCAP",
    volume = "02",
    pages = "020",
    year = "2012"
}

@article{Grosse-Knetter:1992tbp,
    author = "Grosse-Knetter, Carsten and Kogerler, Reinhart",
    title = "{Unitary gauge, Stuckelberg formalism and gauge invariant models for effective lagrangians}",
    eprint = "hep-ph/9212268",
    archivePrefix = "arXiv",
    reportNumber = "BI-TP-92-56",
    doi = "10.1103/PhysRevD.48.2865",
    journal = "Phys. Rev. D",
    volume = "48",
    pages = "2865--2876",
    year = "1993"
}

@article{Burgess:2010zq,
    author = "Burgess, C. P. and Lee, Hyun Min and Trott, Michael",
    title = "{Comment on Higgs Inflation and Naturalness}",
    eprint = "1002.2730",
    archivePrefix = "arXiv",
    primaryClass = "hep-ph",
    reportNumber = "CERN-PH-TH-2010-033, PI-PARTPHYS-174, CERN---PH---TH--2010-033",
    doi = "10.1007/JHEP07(2010)007",
    journal = "JHEP",
    volume = "07",
    pages = "007",
    year = "2010"
}

@article{Karananas:2022byw,
    author = "Karananas, Georgios K. and Shaposhnikov, Mikhail and Zell, Sebastian",
    title = "{Field redefinitions, perturbative unitarity and Higgs inflation}",
    eprint = "2203.09534",
    archivePrefix = "arXiv",
    primaryClass = "hep-ph",
    reportNumber = "LMU-ASC 12/22",
    doi = "10.1007/JHEP06(2022)132",
    journal = "JHEP",
    volume = "06",
    pages = "132",
    year = "2022"
}

@article{Giudice:2024tcp,
    author = "Giudice, Gian F. and Lee, Hyun Min and Pomarol, Alex and Shakya, Bibhushan",
    title = "{Nonthermal heavy dark matter from a first-order phase transition}",
    eprint = "2403.03252",
    archivePrefix = "arXiv",
    primaryClass = "hep-ph",
    reportNumber = "CERN-TH-2024-031, DESY-24-031",
    doi = "10.1007/JHEP12(2024)190",
    journal = "JHEP",
    volume = "12",
    pages = "190",
    year = "2024"
}

@article{Cheung:2007st,
    author = "Cheung, Clifford and Creminelli, Paolo and Fitzpatrick, A. Liam and Kaplan, Jared and Senatore, Leonardo",
    title = "{The Effective Field Theory of Inflation}",
    eprint = "0709.0293",
    archivePrefix = "arXiv",
    primaryClass = "hep-th",
    reportNumber = "IC-2007-032",
    doi = "10.1088/1126-6708/2008/03/014",
    journal = "JHEP",
    volume = "03",
    pages = "014",
    year = "2008"
}

@article{Gubitosi:2012hu,
    author = "Gubitosi, Giulia and Piazza, Federico and Vernizzi, Filippo",
    title = "{The Effective Field Theory of Dark Energy}",
    eprint = "1210.0201",
    archivePrefix = "arXiv",
    primaryClass = "hep-th",
    doi = "10.1088/1475-7516/2013/02/032",
    journal = "JCAP",
    volume = "02",
    pages = "032",
    year = "2013"
}

@article{Tsujikawa:2014mba,
    author = "Tsujikawa, Shinji",
    editor = "Papantonopoulos, Eleftherios",
    title = "{The effective field theory of inflation/dark energy and the Horndeski theory}",
    eprint = "1404.2684",
    archivePrefix = "arXiv",
    primaryClass = "gr-qc",
    doi = "10.1007/978-3-319-10070-8_4",
    journal = "Lect. Notes Phys.",
    volume = "892",
    pages = "97--136",
    year = "2015"
}

@article{Adams:2006sv,
    author = "Adams, Allan and Arkani-Hamed, Nima and Dubovsky, Sergei and Nicolis, Alberto and Rattazzi, Riccardo",
    title = "{Causality, analyticity and an IR obstruction to UV completion}",
    eprint = "hep-th/0602178",
    archivePrefix = "arXiv",
    reportNumber = "CERN-PH-TH-2006-033, HUTP-06-A0005",
    doi = "10.1088/1126-6708/2006/10/014",
    journal = "JHEP",
    volume = "10",
    pages = "014",
    year = "2006"
}

@article{Babichev:2007dw,
    author = "Babichev, Eugeny and Mukhanov, Viatcheslav and Vikman, Alexander",
    title = "{k-Essence, superluminal propagation, causality and emergent geometry}",
    eprint = "0708.0561",
    archivePrefix = "arXiv",
    primaryClass = "hep-th",
    reportNumber = "LMU-ASC-54-07",
    doi = "10.1088/1126-6708/2008/02/101",
    journal = "JHEP",
    volume = "02",
    pages = "101",
    year = "2008"
}

@article{Hollowood:2015elj,
    author = "Hollowood, Timothy J. and Shore, Graham M.",
    title = "{Causality Violation, Gravitational Shockwaves and UV Completion}",
    eprint = "1512.04952",
    archivePrefix = "arXiv",
    primaryClass = "hep-th",
    doi = "10.1007/JHEP03(2016)129",
    journal = "JHEP",
    volume = "03",
    pages = "129",
    year = "2016"
}

@article{deRham:2020zyh,
    author = "de Rham, Claudia and Tolley, Andrew J.",
    title = "{Causality in curved spacetimes: The speed of light and gravity}",
    eprint = "2007.01847",
    archivePrefix = "arXiv",
    primaryClass = "hep-th",
    reportNumber = "Imperial/TP/2020/CdR/03",
    doi = "10.1103/PhysRevD.102.084048",
    journal = "Phys. Rev. D",
    volume = "102",
    number = "8",
    pages = "084048",
    year = "2020"
}

@article{Creminelli:2022onn,
    author = "Creminelli, Paolo and Janssen, Oliver and Senatore, Leonardo",
    title = "{Positivity bounds on effective field theories with spontaneously broken Lorentz invariance}",
    eprint = "2207.14224",
    archivePrefix = "arXiv",
    primaryClass = "hep-th",
    doi = "10.1007/JHEP09(2022)201",
    journal = "JHEP",
    volume = "09",
    pages = "201",
    year = "2022"
}

@article{CarrilloGonzalez:2023cbf,
    author = "Carrillo Gonz{\'a}lez, Mariana and de Rham, Claudia and Jaitly, Sumer and Pozsgay, Victor and Tokareva, Anna",
    title = "{Positivity-causality competition: a road to ultimate EFT consistency constraints}",
    eprint = "2307.04784",
    archivePrefix = "arXiv",
    primaryClass = "hep-th",
    reportNumber = "Imperial/TP/2023/MC/02",
    doi = "10.1007/JHEP06(2024)146",
    journal = "JHEP",
    volume = "06",
    pages = "146",
    year = "2024"
}

@article{Kaplan:2024qtf,
    author = "Kaplan, David E. and Rajendran, Surjeet and Serra, Francesco",
    title = "{Wrong signs are alright}",
    eprint = "2406.06681",
    archivePrefix = "arXiv",
    primaryClass = "hep-th",
    reportNumber = "FERMILAB-PUB-24-0488-SQMS-V",
    doi = "10.1007/JHEP03(2025)031",
    journal = "JHEP",
    volume = "03",
    pages = "031",
    year = "2025"
}

@article{CarrilloGonzalez:2023emp,
    author = "Carrillo Gonz{\'a}lez, Mariana",
    title = "{Bounds on EFT{\textquoteright}s in an expanding universe}",
    eprint = "2312.07651",
    archivePrefix = "arXiv",
    primaryClass = "hep-th",
    reportNumber = "Imperial/TP/2023/MC/03",
    doi = "10.1103/PhysRevD.109.085008",
    journal = "Phys. Rev. D",
    volume = "109",
    number = "8",
    pages = "085008",
    year = "2024"
}

@article{CarrilloGonzalez:2025fqq,
    author = "Carrillo Gonz{\'a}lez, Mariana and C{\'e}spedes, Sebasti{\'a}n",
    title = "{Causality bounds on the primordial power spectrum}",
    eprint = "2502.19477",
    archivePrefix = "arXiv",
    primaryClass = "hep-th",
    reportNumber = "Imperial--TP--2025--SCC--2",
    doi = "10.1088/1475-7516/2025/08/071",
    journal = "JCAP",
    volume = "08",
    pages = "071",
    year = "2025"
}

@article{Hui:2025aja,
    author = "Hui, Lam and Nicolis, Alberto and Podo, Alessandro and Zhou, Shengjia",
    title = "{Microcausality without Lorentz invariance}",
    eprint = "2502.04215",
    archivePrefix = "arXiv",
    primaryClass = "hep-th",
    doi = "10.1007/JHEP07(2025)188",
    journal = "JHEP",
    volume = "07",
    pages = "188",
    year = "2025"
}

@article{Aydemir:2012nz,
    author = "Aydemir, Ufuk and Anber, Mohamed M. and Donoghue, John F.",
    title = "{Self-healing of unitarity in effective field theories and the onset of new physics}",
    eprint = "1203.5153",
    archivePrefix = "arXiv",
    primaryClass = "hep-ph",
    doi = "10.1103/PhysRevD.86.014025",
    journal = "Phys. Rev. D",
    volume = "86",
    pages = "014025",
    year = "2012"
}

@article{Kase:2014cwa,
    author = "Kase, Ryotaro and Tsujikawa, Shinji",
    title = "{Effective field theory approach to modified gravity including Horndeski theory and Ho{\v{r}}ava{\textendash}Lifshitz gravity}",
    eprint = "1409.1984",
    archivePrefix = "arXiv",
    primaryClass = "hep-th",
    doi = "10.1142/S0218271814430081",
    journal = "Int. J. Mod. Phys. D",
    volume = "23",
    number = "13",
    pages = "1443008",
    year = "2015"
}

@article{Kennedy:2017sof,
    author = "Kennedy, Joe and Lombriser, Lucas and Taylor, Andy",
    title = "{Reconstructing Horndeski models from the effective field theory of dark energy}",
    eprint = "1705.09290",
    archivePrefix = "arXiv",
    primaryClass = "gr-qc",
    doi = "10.1103/PhysRevD.96.084051",
    journal = "Phys. Rev. D",
    volume = "96",
    number = "8",
    pages = "084051",
    year = "2017"
}

@article{Aoki:2021wew,
    author = "Aoki, Katsuki and Gorji, Mohammad Ali and Mukohyama, Shinji and Takahashi, Kazufumi",
    title = "{The effective field theory of vector-tensor theories}",
    eprint = "2111.08119",
    archivePrefix = "arXiv",
    primaryClass = "hep-th",
    reportNumber = "YITP-21-132, IPMU21-0079",
    doi = "10.1088/1475-7516/2022/01/059",
    journal = "JCAP",
    volume = "01",
    number = "01",
    pages = "059",
    year = "2022"
}

@article{Mack:2009mi,
    author = "Mack, Gerhard",
    title = "{D-independent representation of Conformal Field Theories in D dimensions via transformation to auxiliary Dual Resonance Models. Scalar amplitudes}",
    eprint = "0907.2407",
    archivePrefix = "arXiv",
    primaryClass = "hep-th",
    month = "7",
    year = "2009"
}

@article{Penedones:2010ue,
    author = "Penedones, Joao",
    title = "{Writing CFT correlation functions as AdS scattering amplitudes}",
    eprint = "1011.1485",
    archivePrefix = "arXiv",
    primaryClass = "hep-th",
    doi = "10.1007/JHEP03(2011)025",
    journal = "JHEP",
    volume = "03",
    pages = "025",
    year = "2011"
}

@article{Ageeva:2021yik,
    author = "Ageeva, Y. and Petrov, P. and Rubakov, V.",
    title = "{Nonsingular cosmological models with strong gravity in the past}",
    eprint = "2104.13412",
    archivePrefix = "arXiv",
    primaryClass = "hep-th",
    reportNumber = "INR-TH-2021-008",
    doi = "10.1103/PhysRevD.104.063530",
    journal = "Phys. Rev. D",
    volume = "104",
    number = "6",
    pages = "063530",
    year = "2021"
}

@article{Marolf:2012kh,
    author = "Marolf, Donald and Morrison, Ian A. and Srednicki, Mark",
    title = "{Perturbative S-matrix for massive scalar fields in global de Sitter space}",
    eprint = "1209.6039",
    archivePrefix = "arXiv",
    primaryClass = "hep-th",
    doi = "10.1088/0264-9381/30/15/155023",
    journal = "Class. Quant. Grav.",
    volume = "30",
    pages = "155023",
    year = "2013"
}

@article{Melville:2023kgd,
    author = "Melville, Scott and Pimentel, Guilherme L.",
    title = "{de Sitter S matrix for the masses}",
    eprint = "2309.07092",
    archivePrefix = "arXiv",
    primaryClass = "hep-th",
    doi = "10.1103/PhysRevD.110.103530",
    journal = "Phys. Rev. D",
    volume = "110",
    number = "10",
    pages = "103530",
    year = "2024"
}

@article{Donath:2024utn,
    author = "Donath, Yaniv and Pajer, Enrico",
    title = "{The in-out formalism for in-in correlators}",
    eprint = "2402.05999",
    archivePrefix = "arXiv",
    primaryClass = "hep-th",
    doi = "10.1007/JHEP07(2024)064",
    journal = "JHEP",
    volume = "07",
    pages = "064",
    year = "2024"
}

@article{Melville:2024ove,
    author = "Melville, Scott and Pimentel, Guilherme L.",
    title = "{A de Sitter S-matrix from amputated cosmological correlators}",
    eprint = "2404.05712",
    archivePrefix = "arXiv",
    primaryClass = "hep-th",
    doi = "10.1007/JHEP08(2024)211",
    journal = "JHEP",
    volume = "08",
    pages = "211",
    year = "2024"
}

@article{DuasoPueyo:2024usw,
    author = "Duaso Pueyo, Carlos and Goodhew, Harry and McCulloch, Ciaran and Pajer, Enrico",
    title = "{Perturbative unitarity bounds from momentum-space entanglement}",
    eprint = "2410.23709",
    archivePrefix = "arXiv",
    primaryClass = "hep-th",
    doi = "10.1007/JHEP08(2025)047",
    journal = "JHEP",
    volume = "08",
    pages = "047",
    year = "2025"
}

@article{Deffayet:2010qz,
    author = "Deffayet, Cedric and Pujolas, Oriol and Sawicki, Ignacy and Vikman, Alexander",
    title = "{Imperfect Dark Energy from Kinetic Gravity Braiding}",
    eprint = "1008.0048",
    archivePrefix = "arXiv",
    primaryClass = "hep-th",
    reportNumber = "CERN-PH-TH-2010-166",
    doi = "10.1088/1475-7516/2010/10/026",
    journal = "JCAP",
    volume = "10",
    pages = "026",
    year = "2010"
}

@misc{donofrio2025validationnanograv15yeardata,
      title={Validation of NANOGrav 15-year data and ACT data by modified inflation in entropic cosmology}, 
      author={Simone D'Onofrio and Sergei Odintsov and Tanmoy Paul},
      year={2025},
      eprint={2510.20484},
      archivePrefix={arXiv},
      primaryClass={gr-qc},
      url={https://arxiv.org/abs/2510.20484}, 
}

\end{document}